\input harvmac.tex
\noblackbox
\input epsf.tex
\overfullrule=0mm
\newcount\figno
\figno=0
\def\fig#1#2#3{
\par\begingroup\parindent=0pt\leftskip=1cm\rightskip=1cm\parindent=0pt 
\baselineskip=11pt
\global\advance\figno by 1
\midinsert
\epsfxsize=#3
\centerline{\epsfbox{#2}}
{\bf Fig. \the\figno:} #1\par
\endinsert\endgroup\par
}
\def\figlabel#1{\xdef#1{\the\figno}}
\def\encadremath#1{\vbox{\hrule\hbox{\vrule\kern8pt\vbox{\kern8pt
\hbox{$\displaystyle #1$}\kern8pt}
\kern8pt\vrule}\hrule}}

\lref\GZ{S. Ghoshal, A. Zamolodchikov, Int. J. Mod. Phys. A9 (1994) 3841.}
\lref\FLSbig{P. Fendley, A. Ludwig, H. Saleur, Phys. Rev. B52 (1995) 8934.}
\lref\BLZ{V. Bazhanov, S. Lukyanov, A.B. Zamolodchikov, Commun.Math.Phys. 
177 (1996) 381-398; Nucl.Phys. B489 (1997) 487-531;
Comm. Math. Phys. 190 (1997) 247.}
\lref\Nat{N. Andrei, K. Furuya, J. Lowenstein, Rev. Mod. Phys. 
55 (1983) 331.}
\lref\PF{P. Fendley, Phys. Rev. Lett. 71 (1993) 2485.}
\lref\GR{I. Gradshtein and I. Rishnik, {\it Table of Integrals, Series and 
Products} (Academic Press, 1980).}
\lref\HMN{P. Hasenfratz, P. Maggiore, F. Niedermayer, 
Phys. Lett. 245 (1990) 522.}
\lref\Zamo{Al. B. Zamolodchikov, Int. J. Mod. Phys. A10 (1995) 1125.}
\lref\FLSbig{P. Fendley, A. Ludwig, H. Saleur,
Phys. Rev. B52 (1995) 8934.}
\lref\sergeiL{S. Lukyanov, ``Low energy effective hamiltonian
for the XXZ spin chain'', cond-mat/9712314.}
\lref\AL{I. Affleck and A. W. W. Ludwig, Phys. Rev.  Lett.
 {\bf 67} (1991) 161.}
\lref\FS{P. Fendley, H. Saleur,  "Massless integrable quantum field
theories and massless scattering in $1+1$ dimensions", Proceedings on
the Trisete Summer School in High Energy Physics and Cosmology,
(1993), Gava et al. Eds., World Scientific}
\lref\Zamoprivate{Al. B. Zamolodchikov, private communication in 1996}
\lref\Hew{A. C. Hewson, ``The Kondo problem to heavy fermions'', Cambridge
University Press (1993).}
\lref\Gh{S. Ghoshal, Int. J. Mod. Phys. A9 (1994) 4801.}
\lref\KF{C. Kane, M.P.A. Fisher, Phys. Rev. B46 (1992) 15233.}
\lref\sch{Albert Schmid, Phys. Rev. Lett. 51 (1983) 1506.}
\lref\LSshort{F. Lesage, H. Saleur,
``Strong coupling resistivity in the Kondo model'', cond-mat/9811172.}
\lref\miscel{P.P. Kulish, E.R. Nisimov, Th. and Math. Phys.
29 (1976) 992; M. L\"uscher, Nucl. Phys. B117 (1997) 475; 
T. Eguchi and S. K. Yang, Phys. Lett. B224 (1989) 373; R. Sasaki, I. Yamanaka,
Adv. Studies in Pure Math. 16 (1988) 271.}
\lref\ABL{C. Ahn, D. Bernard, A. Leclair, Nucl. Phys. B346 (1990) 409.}
\lref\Claudio{C. De Chamon, D. Freed, X.G. Wen, Phys. Rev. B51 (1995) 2363.}
\lref\sergei{S. Skorik, `` Exact non-equilibrium current from the partition function for impurity transport
problems,
cond-mat/9707307.}
\lref\FLeS{P. Fendley, F. Lesage, H. Saleur, J. Stat. Phys. 85 (1996) 211.}
\lref\TW{A.M. Tsvelick and P. B. Wiegmann, Z. Phys. B54 (1985) 201; J. Stat. Phys. 38 (1985) 125; A.M. Tsvelick,
J. Phys. C18 (1985) 159.}
\lref\affleck{I. Affleck, A.W.W.
Ludwig, Phys. Rev. B48 (1993) 7297.}
\lref\cost{T.A. Costi, A.C. Hewson, Phil. Mag. B65, (1992) 1165;
T.A. Costi, Private communication.}
\lref\AGD{A.A. Abrikosov, L.P. Gorkov and I.E.Dzyaloshinski, ``Methods
of quantum field theory in statistical physics'', Dover, New York (1963).}
\lref\FL{F. Lesage, H. Saleur, Nucl. Phys. B490 [FS], (1997) 543.}
\lref\pauldual{P. Fendley, ``Duality without supersymmetry'', hep-th/9804108.}
\lref\usdual{P. Fendley, H. Saleur, Phys. Rev. Lett. 81 (1998) 2518; 
``Hyperelliptic curves for multi-channel quantum wires and the multi-channel
Kondo problem'', cond-mat/9809259.}

\Title{USC-98-009}
{\vbox{
\centerline{Perturbation of infra-red fixed points and duality}
 \vskip 4pt
\centerline{in quantum impurity problems.}}}

\centerline{F. Lesage$^*$, H. Saleur$^{**}$}
\bigskip\centerline{$^*$Centre de recherches math\'ematiques,}
\centerline{Universit\'e de Montr\'eal,}
\centerline{C.P. 6128 Succ. centre-ville, Montr\'eal,  H3C-3J7.}

\bigskip\centerline{$^{**}$Department of Physics}
\centerline{University of Southern California}
\centerline{Los Angeles, CA 90089-0484}

\vskip .3in

We explain in this paper how
a meaningful irrelevant perturbation theory around the infra-red (strong coupling) 
fixed point can be carried out
for integrable quantum impurity problems.
This is illustrated in details for the spin $1/2$ Kondo model,
where our approach gives rise to the  complete low temperature expansion of the resistivity,
beyond the well known $T^2$ Fermi liquid behaviour. We also consider 
the edge states tunneling problem, and demonstrate by Keldysh techniques 
that the DC current satisfies an {\sl exact duality} between the UV and IR regimes. This 
corresponds physically to a duality between the tunneling of 
Laughlin quasi particles and electrons, and, more formally, to the existence of an exact instantons expansion. 
The duality is deeply  
connected with integrability, and could not have been expected a priori.

\Date{11/98}

\newsec{Introduction}

Duality arguments have been commonly used in quantum impurity problems for many years. An archetypal
situation is provided by the model of a particle moving in a periodic
potential and subject to quantum dissipation
\sch. This 
problem is represented in one dimension \foot{Space dimensionality does not play a crucial role here}
 by the following (boundary sine-Gordon or BSG) hamiltonian
\eqn\bdrsg{{\cal H}={1\over 2}\int_{-\infty}^0dx\int_{-\infty}^\infty
dy \left[(\partial_x\Phi)^2+(\Pi)^2\right]+2\lambda\cos\sqrt{2\pi g}\Phi(0).}
where the field at the origin represents the particle coordinate, and the bulk 
free boson the bath degrees of freedom. At small $\lambda$ the particle diffuses freely (UV fixed point)
, while
at large $\lambda$, it is localized (IR fixed point) in a minimum of the potential, given by $\Phi(0)=\sqrt{2\pi\over g}
n$, 
$n$ an integer. Near the UV fixed point, physical properties can be expanded 
in powers of $\lambda$, and are expressed in terms of Coulomb gas integrals
whose charges $\pm g$ correspond to the two possible exponentials in the cosine term (the dimension of the 
operator $\cos\sqrt{2\pi g}\Phi(0)$ being $\Delta=g$).
It is also possible to study the vicinity of this IR fixed point in an ${1\over\lambda}$ expansion 
by considering the instantons and anti-instantons  that take the particle from one minima to a neighbouring one.
Using  the leading order  action of these instantons, one obtains again a Coulomb gas, but this time 
with charges $\pm {1\over g}$. This demonstrates, in slightly more formal terms,  that the leading IR hamiltonian
looks as \bdrsg, but with a perturbation $\lambda_d\cos\sqrt{2\pi\over g}\tilde{\Phi}$ (with
dimension $\Delta_d={1\over g}$), where $\tilde{\Phi}$
is the dual of the free boson in the usual sense, and by dimensional analysis, $\lambda_d\propto
 \lambda^{-{1\over g}}$.

The same hamiltonian \bdrsg\ appears also in the problem of tunneling between edge states in the 
fractional quantum Hall effect \KF. In that case, while it is Laughlin quasi particles of charge $g=\nu$ (the filling
fraction) that tunnel in the UV, the duality argument demonstrates that it is 
electrons of charge unity that tunnel in the IR. 

The duality just discussed is very useful qualitatively. It has however been used in the literature
as a much stronger statement: namely that physical properties should exhibit an {\sl exact} duality
between the UV and IR fixed points under replacement of $\lambda$ by $\lambda_d$ and $g$ by ${1\over g}$. 
Why this should be the case was not explained, and it  must be stressed that this is a highly
non trivial result: the approach to the IR (strong coupling) 
fixed point is, in general, determined by a 
very specific combination of irrelevant operators coming with amplitudes that are all
powers of $\lambda_d$, so they all contribute equally  significantly: for instance, one expects
that, in addition to the term $\lambda_d\cos\sqrt{2\pi\over g}\tilde{\Phi}$,  terms
$\lambda_d^{n^2\Delta_d-1\over \Delta_d-1
}\cos n\sqrt{2\pi\over g}\tilde{\Phi}$ should also appear (where $\Delta_d={1\over g}$), corresponding 
physically to multi-instantons processes, or tunneling of several electrons. Such terms might also be
 required as counterterms to cure the very strong short distance divergences of the IR perturbation theory.
Clearly, the existence of these terms will destroy any hope of observing an exact duality, and one should not
expect the duality argument to tell us more than the leading irrelevant operator, in a general situation.

Nevertheless, an analytical  computation of the mobility (the current) at $T=0$ and with an external
force (bias) \FLSbig\ has exhibited an exact duality between the UV and IR for the model \bdrsg, adding up confusion
to the whole issue. Something 
very special must be happening in that case - and indeed, the model is integrable. 

One of the purposes of this paper is to discuss why integrability gives rise to an exact duality
for some physical properties - and also, to explain why this duality should not 
be expected for other properties. In discussing these questions, 
we will actually consider   IR perturbation theory, and show how it can  be made
meaningful,  again thanks to integrability. This has applications beyond the tunneling 
problem: as an example, we discuss  in details the case of the resistivity in the Kondo problem.

In the second section of this paper, we use the simple example of the Ising model
with a boundary magnetic field to discuss how the integrable structure of quantum impurity problems
gives a quick access to the full hamiltonian near the IR fixed point, which is essentially 
encoded in the reflection matrix, or the boundary free energy. We discuss the issue of regularization
for the IR perturbation theory that arises in integrable models, and how one can use the renormalization
group backwards in some cases. 

In the third section of this paper, we discuss how the IR action can be determined for 
the spin $1/2$ Kondo problem and for the tunneling problem. We also sketch the 
result for the higher spin Kondo problem.

In the fourth section of this paper, we discuss, as an application of  IR 
perturbation theory, the resistivity in the Kondo problem -  and determine, 
beyond the well known $T^2$ order, its complete  low temperature behaviour.

In the fifth section of this paper, we finally discuss duality issues. We show that
the anisotropic higher spin Kondo model exhibits
a partial duality, manifest for instance in the following relation 
\eqn\maini{f\left(j,\lambda,H,g\right)\equiv f\left(j-{1\over 2},\lambda_d,{H\over g},{1\over g}\right),}
that holds up to analytical terms (odd powers) in $H/T_B$. We also show that 
the  current in the tunneling problem
obeys an exact duality  
\eqn\mainii{I(\lambda,g,V,T)=gV-gI\left(\lambda_d,{1\over g},gV,T\right).}

These duality properties follow 
from the structure of the IR hamiltonians that is strongly constrained by 
integrability: within our ``analytic'' regularization scheme, they
   are made up of  an infinite series of 
local (conserved) quantities (polynomials in derivatives of $\Phi$), 
plus at most {\sl one} non local term, which is $\lambda_d
\cos\sqrt{2\pi\over g}\tilde{\Phi}(0)$
for the BSG case, and $\lambda_dS_-e^{i\sqrt{2\pi\over g}\tilde{\Phi}(0)}+cc$ for the  spin $j$
Kondo case (where here $S_\pm$ are spin $j-1/2$ operators) (all this within a well defined
 regularization scheme). As a result, thermodynamic quantities
will in general exhibit a partial duality; the UV expansion in  even powers of $\lambda$ will match the
part of the  IR expansion
that is in even powers of $\lambda_d$, although there will also be other terms in this IR expansion
due to the local conserved quantities. Some other properties turn out to be blind to 
the local conserved quantities however, and as a result exhibit an {\sl exact} duality,
like the DC current in the tunneling problem.

In the first appendix, we determine
the normalization of conserved quantities in the sine-Gordon theory. 
The second appendix contains some remarks about the Keldysh formalism and analytic continuation.

Some of the results presented here have appeared in short form  in \LSshort. The methods we develop
are related, although independent and different, to the series of works by Bazhanov, Lukyanov and
Zamolodchikov  \BLZ, and also to the work of Lukyanov \sergeiL. Duality in quantum impurity problems
has also been investigated by Fendley  \pauldual, and by Fendley and  one of us \usdual,
from a more formal perspective.

\newsec{Getting the IR hamiltonian in integrable boundary field theories: the case of the Ising model.}

\subsec{Some generalities}

We consider the Ising model defined on the half space 
$x\in [-\infty,0]$, $y\in
[-\infty,\infty]$. We initially 
use a crossed channel or  {\sl open string} description, where euclidian time 
runs in the $y$ direction, and  we introduce 
the complex coordinate $w=-y+ix$.  We restrict to the case
of a theory that is massless in the bulk, and add up
a boundary magnetic field $h$ (Fig. 1). 
\fig{Geometry of the problem.}{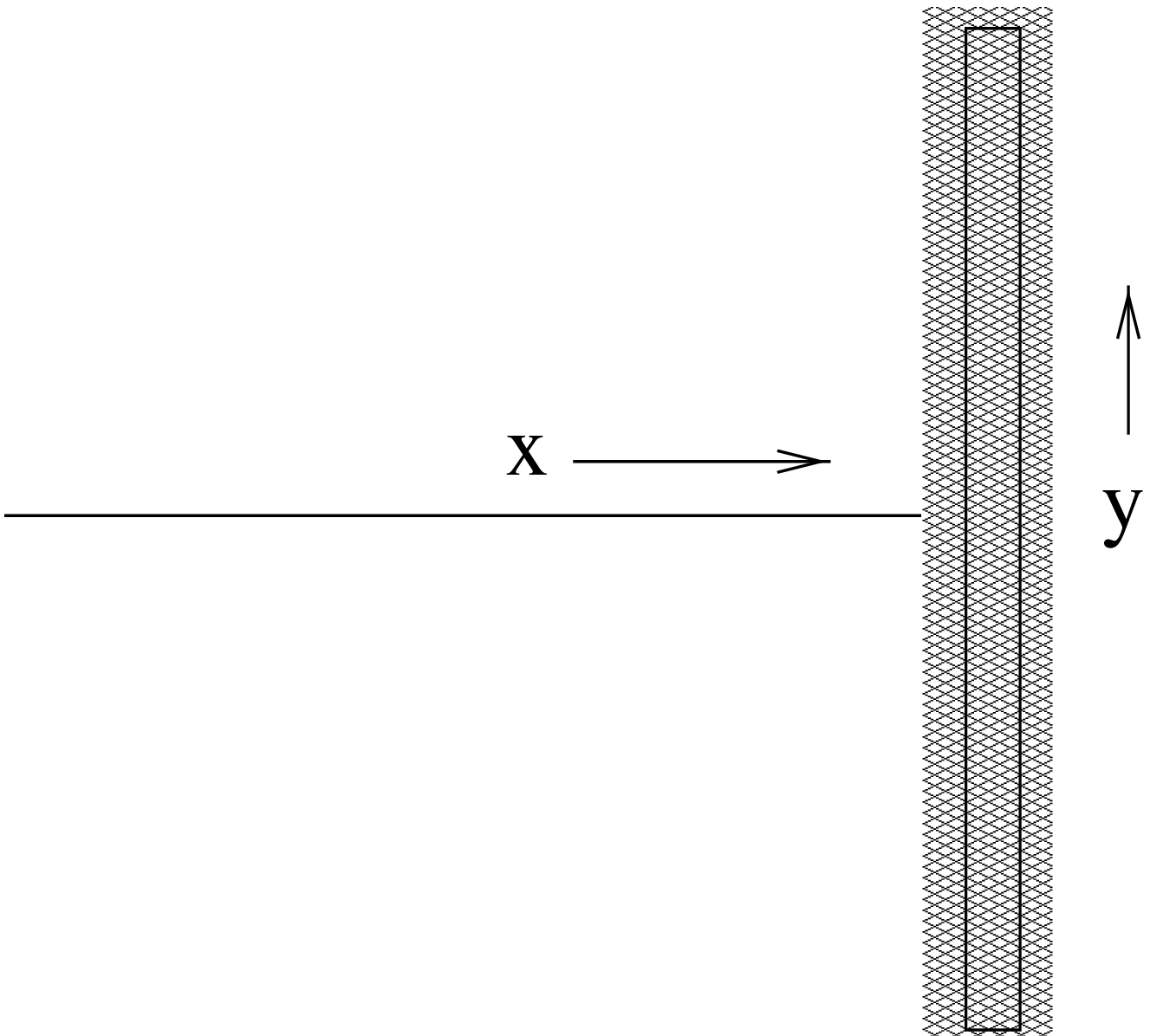}{6.5cm}
\figlabel\tabb

When $h=0$, the fermions have 
free boundary conditions $\psi_L=\psi_R$ on the
boundary; this is the UV fixed point. When $h\to\infty$, the Ising spins 
become fixed, corresponding to $\psi_L=-\psi_R$; this is the IR fixed point. 
The question we wish to study is how the IR fixed point is approached. More 
precisely maybe, we want to be able to describe the Ising model at large values of 
$h$ with a hamiltonian
$$
{\cal H}={\cal H}_{IR}+\delta{\cal H}(0)
$$
where ${\cal H}$ expands in some  powers of the inverse coupling constant $1/h$.
It is possible to gain some quick insight on what $\delta{\cal H}$ should
 look like.  - 
 it is actually an expansion in odd powers of $1/h^2$. This is
 because the operator content of the Ising model with fixed boundary conditions
can easily be extracted from conformal invariance considerations: 
with fixed boundary conditions on a cylinder of length $L$ and circumference $1/T$,
 the Ising partition function
is simply the identity character (setting $q=e^{-\pi/LT}$) 
$$
\chi_0={1\over 2}\left[\prod_0^\infty \left(1-q^{n+1/2}\right)+  
\prod_0^\infty \left(1
+q^{n+1/2}\right)\right].
$$
{}From this, it follows that the only available 
 operators are of the form  $\partial^p\psi_R\partial^q\psi_R+(R\to L)$ (here,
$\partial$ stands for $\partial_w$). Up to 
total derivatives which do not affect the physical properties 
of interest, we can restrict  to $\psi_R\partial^n\psi_R+(R\to L)$,
with $n$ odd. 
Introducing the operator (the normalization is chosen
for later convenience)
\eqn\iconsquant{{\cal O}_{2k+2}^o=
(-1)^{k+1}{1\over 4}\left(:\psi_R\partial_w^{2k+1}\psi_R:+:
\psi_L\partial_{\bar{w}}^{2k+1}\psi_L:\right),}
we thus expect 
\eqn\genestructure{
\delta{\cal H}=\sum_{k=0}^\infty  a_{2k+1}{1\over h^{2(2k+1)}}{\cal O}^o_{2k+2}(0),}
where the coefficients $a_{2k+1}$ have to be determined. Notice that in practice,
the manipulation of expressions like \genestructure\  will give rise to
extremely strong short distance divergences - the numerical
values of $a_{2k+1}$ will only have a well defined meaning 
within a specific regularization scheme.

For a general problem, such a computation would appear
untractable. What makes it feasible in the cases we are going to consider
in this paper is integrability. To see how this comes about, 
and pave the way for generalizations, let us   describe the Ising model
using massless scattering. In this simple case, we have massless 
R and L moving fermionic particles with 
 energy and momentum parametrized as  $e=\pm p= e^\beta$, $\beta$ the 
rapidity.  
The mode expansion of the fermion operators is 
\eqn\fermexp{\eqalign{\psi_R(w)=\int {d\beta\over 2\pi} e^{\beta/2}
\left[\omega\exp(e^\beta w) Z_R(\beta)+\bar{\omega}\exp(-e^\beta
w)Z_R^*(\beta)\right]\cr
\psi_L(\bar{w})=\int {d\beta\over 2\pi} e^{\beta/2}
\left[\bar{\omega}\exp(e^\beta \bar{w}) Z_L(\beta)+\omega\exp(-e^\beta
\bar{w})Z_L^*(\beta)\right],\cr}}
where the $Z$ are  creation and annihilation
operators obeying the usual anticommuting relations, $\omega=e^{i\pi/4}$. The theory
is defined on the
half space $x\in [-\infty,0]$ only; as a result, the L and R modes are
not independent. 
Because the boundary interaction is integrable, the fermions, 
in the crossed channel picture scatter off the boundary one by one, with no particle
production, and one has $Z^*_R(\beta)=R(\beta)Z^*_L(\beta)$, $R$ the 
reflection matrix \GZ.

 We will also use the 
 direct channel or {\sl closed string} picture, where euclidian time runs
in the $x$ direction.  The mode expansion 
of the fermions is identical  to \fermexp, with $w$ replaced by 
the variable $z=w/i=x+iy$. 
 The Hilbert space 
is then the usual one for fermions defined on the whole line, and there is 
no relation between L and R modes. Rather,  
in the direct channel picture, the effect of the boundary
 is taken into 
account by the existence of a {\sl boundary state}, which reads \GZ\ \foot{We do not discuss
the problem of the overall normalization of the boundary state in this multiparticle
description - it is enough to recall that it is independent of $\beta_B$.}
\eqn\bdrstate{
|B\rangle \propto\sum_{n=0}^\infty {1\over n!}\int
\prod_i {d\beta_i\over 2\pi}
K(\beta_i-\beta_B)
Z_L^*(\beta_i)Z^*_R(\beta_i)|0\rangle,}
with $K(\beta)=R\left({i\pi\over 2}-\beta\right)$. In the 
simple case of a boundary magnetic field considered so far, 
$K(\beta-\beta_B)=i\tanh{\beta_B-\beta\over 2}$.
The parameter $\beta_B$ is in general related with a typical energy scale
associated with the boundary interaction, $T_B= e^{\beta_B}$. In the 
case of a boundary magnetic field, $T_B\propto h^2$. In the closed string channel,
we introduce the equivalent of \iconsquant\ 
\eqn\iconsquantdir{{\cal O}_{2k+2}=
{1\over 4}\left(:\psi_R\partial_z^{2k+1}\psi_R:+:
\psi_L\partial_{\bar{z}}^{2k+1}\psi_L:\right).}

\subsec{The complete  IR action}

Let us now discuss how the IR action can be simply extracted from the knowledge of the reflection matrix, or,
equivalently, of the boundary state.

To do so, let us keep working in the closed string channel, and consider  the expression for 
the boundary state $|B>$ further. The IR boundary state (fixed 
boundary conditions) is obtained as $\beta_B\to\infty$
where $K=i$:
$$
|B_{IR}\rangle\propto \sum_{n=0}^\infty {i^n\over n!}\int
\prod_i{d\beta_i\over 2\pi}
Z_L^*(\beta_i)Z^*_R(\beta_i)|0>.
$$
One can thus write  $|B\rangle={\cal B}|B_{IR}\rangle$, where the operator ${\cal B}$ 
is defined in the multiparticle basis by
$$
{\cal B}\prod_i Z_L^*(\beta_i)Z^*_R(\beta_i)|0>=
\prod_i {K(\beta_i-\beta_B)\over i}Z_L^*(\beta_i)Z^*_R(\beta_i)|0>.
$$
Let us expand
\eqn\expad{\ln\left[{K(\beta-\beta_B)\over i}\right]=\sum_{k=0}^\infty 
{-2\over 2k+1}\ 
e^{(2k+1)(\beta-\beta_B)}.}
Introduce then the set of {\sl commuting} 
operators ${\cal I}_{2k+1}$ acting on the
multiparticle states,
with 
\eqn\multipartope{{\cal I}_{2k+1}|\beta_1\ldots\beta_n>_{C_1,\ldots,C_n}
={1\over 2}\left(\sum_i
e^{(2k+1)\beta_i}
\right)
|\beta_1\ldots\beta_n>_{C_1,\ldots,C_n},}
where $C=L,R$ designates the chirality. One can then write 
\eqn\main{|B>=\exp\left[\sum_{k=0}^\infty {-2\over 2k+1}\ e^{-(2k+1)\beta_B}
{\cal I}_{2k+1}\right]|B_{IR}\rangle.}
Of course, the ${\cal I}_{2k+1}$ can be expressed in terms of the
creation/annihilation operators, ${\cal I}_{2k+1}=\int {d\beta\over
4\pi}
e^{(2k+1)\beta}\left[Z_L^*(\beta)Z_L(\beta)+Z^*_R(\beta)
Z_R(\beta)\right]$.
Using the mode expansion of the fermions, one checks this 
coincides with $\int_{-\infty}^\infty dy {\cal O}_{2k+2}$,
where ${\cal O}_{2k+2}$ is defined in \iconsquantdir.

We can now write a reasonable conjecture for the hamiltonian
\genestructure\  in the crossed channel - the reason why it is a conjecture only
is because the exponential in \main\ is determined by the action on one 
particular state only, $|{\cal B}_{IR}\rangle$, and not in true generality (one can determine
the action of the exponential on other states with ``momentum'' actually, but still,
not on all possible states of the theory). Observe now that 
if ${\cal H}={\cal H}_{IR}+\delta{\cal H}$,
the boundary state will generally read
 $|B\rangle={\cal P} \exp\left[-\int_{-\infty} ^\infty dy\delta{\cal H}\right]
|B_{IR}\rangle$, where ${\cal P}$ 
is the ($y$) path ordered exponential. Using that  the ${\cal I}_{2k+1}$ form a set of 
commuting quantities, together with the fact that the ${\cal O}_{2k+2}$ 
are self and mutually local operators,  we obtain therefore 
\eqn\action{{\cal H}={\cal H}_{IR}+
\sum_{k=0}^\infty {2\over 2k+1} e^{-(2k+1)\beta_B}
{\cal O}^o_{2k+2}(0),}
again up to total derivatives. 

Because the ${\cal I}_{2k+1}$ form a set of commuting quantities,
the perturbation of the IR hamiltonian in \action\ is, formally, integrable. This is 
an expected result, since after all the flow from the UV to the IR fixed point {\sl is}
integrable, a feature that should be observed from both extremities - and provides an
immediate check of \action.

\subsec{The boundary free energy}

We now discuss the relation between the IR hamiltonian, and the boundary free energy.

Consider thus the theory defined for $x\in [-\infty,0]$ and $y\in [0,1/T]$,
with periodic boundary conditions in the $y$ direction. In the 
closed string  point of view, the theory is thus defined on a circle
instead of the infinite line, while in the open string
point of view, it is now at finite temperature $T$. 

To compute the free energy, it is convenient to ``unfold'' the problem,
so now $x\in [-\infty,\infty]$, and the boundary interaction 
becomes an ``impurity interaction'' acting only on the R movers. Notice
that a different  unfolding is appropriate 
to study the vicinity of the UV and the IR
fixed point; in one case, one extends the theory to $x>0$ by
setting $\psi_R(x,y)=\psi_L(-x,y)$, while in the other one sets of
course $\psi_R(x,y)=-\psi_L(-x,y)$. In what follows, we discuss
mostly the vicinity of the IR fixed point, and thus use the second folding. 
{}From the resulting ``impurity'' point of view, the IR fixed point is 
then just described by free R moving fermions.

Introducing
 then, in the closed string channel
\eqn\newi{{\cal I}_{2k+1}=\int_0^{1/T} {\cal O}_{2k+2}(z) dy=
{1\over 2}\int_0^{1/T} dy :\psi_R\partial_z^{2k+1}
\psi_R:,}
it follows from the expression of the boundary state
that \foot{Here we have subtracted all extensive non universal terms.}
\eqn\partfct{f=-T\ln g_{IR}-T\sum_{k=0}^\infty {-2\over 2k+1}
e^{-(2k+1)\beta_B} 
{}_{1/T}\langle 0|{\cal I}_{2k+1}|0\rangle_{1/T},}
where $g_{IR}={}_{1/T}\langle 0|B_{IR}\rangle_{1/T}$ is the boundary degeneracy of the
IR boundary state (actually independent of $T$ \AL)
and $|0\rangle_{1/T}$ denotes the ground state of the theory
on a circle of circumference $1/T$. 

The ground state on a circle corresponds to fermions
with antiperiodic boundary conditions. Using the mode 
expansion of the fermions, and the expression
of   ${\cal I}_{2k+1}$ as the sum of the $2k+1^{th}$ powers of the energy, it 
follows that\foot{We use here the well known fact that the ``conformal normal ordering''
is related to the ``operator normal ordering'' by zeta regularization of the divergent parts.} 
\eqn\ef{\eqalign{{}_{1/T}\langle 0|{\cal I}_{2k+1}|0\rangle_{1/T}=&{1\over 2}
\left(
2\pi T\right)^{2k+1}
\langle 0|\sum_{j=-\infty}^{\infty}(j+1/2)^{2k+1}\psi_{-j-1/2}\psi_{j+1/2}
|0\rangle\cr &=-{1\over 2}\left(2\pi T\right)^{2k+1}
\sum_{j=0}^ {\infty}(j+1/2)^{2k+1}.\cr}}

The sum can be evaluated by $\zeta$-function regularization leading to 
\eqn\efI{{}_{1/T}\langle 0|{\cal I}_{2k+1}|0\rangle_{1/T}
={1\over 2}\left(2\pi T\right)^{2k+1}\left
(1-{1\over 2^{2k+1}}\right)\zeta(-2k-1),}
(the same computation would give a vanishing 
result for even powers due to  $\zeta(-2k)=0$). For $k=0$ one gets
$-{\pi cT\over 12}$ with $c=1/2$; this is because ${\cal I}_1$ is nothing 
but the zero mode of the stress energy tensor \foot{For the Ising 
model considered here, $T_{zz}=\pi :\psi_R\partial\psi_R:$}
 on a circle, ${\cal I}_1=(2\pi T)
\left(L_0-{c\over 24}\right)$. By plugging the results \efI\ back in
\partfct,  one obtains an explicit expression for $f$. 

Of course, the computation can be done in the open string channel as well;
the expression \action\ for the hamiltonian leads to 
\eqn\crossedexp{f=-T\ln g_{IR}+\sum_{k=0}^\infty {2\over 2k+1}e^{-(2k+1)\beta_B} 
\langle {\cal O}^o_{2k+2}\rangle_T.}
Here, $\langle{\cal O}^o_{2k+2}\rangle_T$ follows from expression \iconsquantdir\  
evaluated in 
the multiparticle basis of the open channel:
\eqn\newj{\langle{\cal O}^o_{2k+2}\rangle_T= (-1)^{k+1}{1\over L}\int {d\beta\over 2\pi}
e^{(2k+1)\beta}\langle Z^*_R(\beta)Z_R(\beta)\rangle_T,}
and the notation $\langle .\rangle_T$ designates the thermal
average in the theory at temperature $T$. 

A standard thermodynamic analysis gives
\eqn\epf{\eqalign{\langle {\cal O}^o_{2k+2}\rangle_T&=(-1)^{k+1} {1\over L}
\int_{-\infty}^\infty{d\beta}
 e^{(2k+1)\beta}\rho(\beta)\cr
&=(-1)^{k+1} {1\over L}\int_{-\infty}^\infty{d\beta}
e^{(2k+1)\beta}  (\rho(\beta)+\tilde\rho(\beta))
{1\over 1+\exp(\epsilon/T)}\cr
&=(-1)^{k+1} \int_{-\infty}^\infty{d\beta\over 2\pi}
  e^{(2k+1)\beta}{d\epsilon\over d\beta}
{1\over 1+\exp(\epsilon/T)},\cr
}}
where, for free fermions, $\epsilon=e^\beta$, and thus 
$2\pi(\rho+\tilde\rho)= L{d\epsilon\over d\beta}$, a result that generalizes 
to interacting theories. By expanding the filling fraction, one obtains
\eqn\epfII{\langle {\cal O}^o_{2k+2}\rangle_T=(-1)^{k+1}
{(2k+1)!\over 2\pi}T^{2k+2}\left
(1-{1\over 2^{2k+1}}\right)\zeta(2k+2).}
To compare \efI\ and \epfII\ recall the identities
\GR\
$$\zeta(2k+2)={(2\pi)^{2k+2}\over 2 (2k+2)!}(-1)^k B_{2k+2};\qquad
\zeta(-2k-1)=-{B_{2k+2}\over 2k+2},$$
where $B_{n}$ are Bernouilli numbers. Hence, as of course should be,
$\langle {\cal O}^o_{2k+2}\rangle_T= T{}_{1/T}\langle 0|{\cal I}_{2k+2}|0\rangle_{1/T}$,
and we find the same expression 
for the impurity free energy \FS. 

Using the thermodynamic expression for the integrals of motion (the first
equation of \epf), we obtain an alternate formula for the 
impurity free energy:
\eqn\tbaimpfr{\eqalign{
f=&-T\ln g_{IR}+\int {d\beta\over 2\pi} \sum_{k=0}^\infty 
{2\over 2k+1}\ e^{(2k+1)(\beta-\beta_B)}(-1)^{k+1} {d\epsilon\over d\beta}
{1\over 1+e^{\epsilon(\beta)/T}}\cr
=&-T\ln g_{UV}-T\int {d\beta\over 2\pi}{1\over\cosh(\beta-\beta_B)}
\ln\left(1+e^{-\epsilon(\beta)/T}\right),\cr}}
where we used the fact that $f\approx -T\ln g_{UV}$ (resp. $f\approx -T\ln g_{IR}$)
as $\beta_B\to -\infty$ (resp. $\beta_B\to\infty$).
This last expression coincides with a well known formula obtained using the 
thermodynamic Bethe ansatz. It reads as well
\eqn\shortcut{f=-T\ln g_{UV}-T\int {d\beta\over 2\pi}{1\over i}{d\over d\beta}\ln R(\beta-\beta_B)
\ln\left(1+e^{-\epsilon(\beta)/T}\right),}
a result that follows directly from the form of the boundary state, and the manipulations
in \epf. 

In the foregoing paragraphs, we have thus showed how the IR action could be 
extracted from the R matrix, and how it was closely related with the boundary free energy. 

\subsec{Flowing ``back'' from the IR fixed point.}

The previous analysis shows very clearly how the TBA results are directly
connected with an IR  description of the flow; in fact,
the free energy provides an immediate reading of the {\sl complete IR action}
(that the impurity free energy  has to do with conserved quantities 
was observed in the earlier papers on the subject already, see \Nat).
It is important to realize that all this works for a particular {\sl regularization}
scheme, involving  dimensional regularization and (or) contour deformation. This is
 somewhat 
obvious since the integrable approach does not 
involve any length scale that could act as a cut-off. The 
 quickest way to
see this more explicitely is to consider for instance the quantity ${\cal I}_1$. Using the 
mode expansion, $|0\rangle_{1/T}$ is clearly 
 an {\sl eigenstate} of ${\cal I}_1$, and thus 
\eqn\discu{
{}_{1/T}\langle 0|\left({\cal I}_1\right)^{p}|0\rangle_{1/T}=
\left({}_{1/T}\langle 0|{\cal I}_1|0\rangle_{1/T}\right)^p=
\left(-{\pi T\over 24}\right)^p.}
To write \discu, we have used an operator formalism, which, in fact requires 
``time ordering'' - here ordering along x. In other words, in \discu, the divergences
 have been
regulated by slightly displacing the $p$ contours of integration. 

The effect of this displacement can be seen
 by using the  fermion propagators
 and Wick's theorem.
Aside from the term involving the (non vanishing)
 average of ${\cal I}_1$ on the circle, contractions  contribute integrals 
with strong short distance divergences, the simplest one being
$$
\int_0^{1/T} dy_1dy_{2}{1\over \left[\sin \pi T(y_1-y_2)\right]^4}.
$$
If one evaluates
this integral by displacing the contours and using the residue theorem (together
with the periodicity of the integrand), one finds indeed a vanishing result, 
because   the integrand has a vanishing residue at the origin.

Equivalently, in dimensional regularization, one considers the more general integral
where the power is a number $\alpha$ ($\alpha=4$ here) 
and one computes the integral in the domain of $\alpha$'s where it is defined. This
gives
$$
\pi^2T^2 {\pi 2^{1+\alpha}\Gamma(2-\alpha)\over (1-\alpha)
\Gamma^2(1-{\alpha\over 2})}
$$
One then continues analytically to $\alpha=4$ - and  the last expression vanishes
again, this time due to the double pole (in $\alpha$) in the denominator. This
generalizes to all the other integrals, so that the dimensionally 
regularized value of ${\cal I}_1^p$ is contributed only 
by its average, ie the result \discu.

In general, integrals involving local conserved quantities (all the ones in the Ising model
are of that type) will be regulated by operator methods or contour displacement, since there is 
no readily available parameter to perform continuations (the prescription is not ambiguous thanks to the commutativity 
of the conserved quantities). Integrals involving 
non local conserved quantities will be regulated by continuation in the parameter $g$. We refer
to this scheme as an ``analytic'' regularization.

Using the previous ideas, it is clear that we can solve the problem of the most general
perturbation of the IR fixed point. For an action
\eqn\mostgene{{\cal H}={\cal H}_{IR}+\sum_{k=0}^\infty b_{2k+1}
{\cal O}^o_{2k+2},}
the boundary free energy simply reads
\eqn\bdrgene{f=-T\ln g_{IR}+\int {d\beta\over 2\pi}\sum_{k=0}^\infty b_{2k+1}
e^{(2k+1)\beta} (-1)^{k+1}{d\epsilon\over d\beta}{1\over 1+e^{\epsilon(\beta)/T}},}
the integrals themshelves being evaluated in \epf,\epfII. 

As the temperature is lowered, ie when one considers this system 
at larger and larger scales, one simply flows to the IR 
fixed point, as physically expected, since all the operators ${\cal O}^o_{2k+2}$
are irrelevant near this fixed point. As the temperature is 
increased, ie when one considers the system at smaller
and smaller scales, or tries to ``flow back'', what happens
generically is that no fixed point is reached; rather, the amplitude
of all the terms becomes bigger and bigger, as expected 
for irrelevant perturbations. The cases where one flows back
to an interesting fixed point are the ones for which
the series in \bdrgene\ defines a function of $T$ which, 
continued beyond the radius of convergence,
has a finite $T\to\infty$ limit. Though we do not 
know any definite mathematical statement about that question, it seems 
clear that these cases are extremely rare. For instance, the choice
 $b_{2k+1}={2\over 2k+1}
e^{-(2k+1)\beta_B}$ guarantes a flow back to the 
free fixed point, but any perturbation that differs, even
infinitesimally,  from this one by a finite
number of terms, will {\sl not} flow back to the free fixed point at all.

A quick way to build an IR hamiltonian that has a $T\to\infty$ limit
is to multiply the reflection matrix by a CDD factor. By flowing backwards,
one finds in this case that the difference $g_{UV}-g_{IR}$ is increased by 
a term $\ln\sqrt{2}$, corresponding presumably to the appearance of additional boundary 
degrees of freedom in the UV.

\newsec{Approach to the IR fixed point for the  spin $1/2$ Kondo model and the boundary sine-Gordon model}

\subsec{The Kondo model}

The previous structure generalizes in a slightly more 
complicated form to the case of the spin $1/2$ Kondo model with action
\eqn\bdrsg{{\cal H}={1\over 2}\int_{-\infty}^0dx\int_{-\infty}^\infty
dy \left[(\partial_x\Phi)^2+(\Pi)^2\right]
+\lambda
\left[S_-e^{i\sqrt{2\pi g}\Phi(0)}+S_+e^{-i\sqrt{2\pi g}\Phi(0)}\right],}
where $S$ is a spin one half operator ($\lambda$ is assumed positive in
what follows). The boundary interaction is
integrable, and the same manipulations we carried out for the Ising model
can be acomplished here too. 

Instead of describing the bulk with massless fermions, we use massless
L and R moving solitons and antisolitons. Parametrizing their energy
by a rapidity $e=\pm p=e^{\beta}$, these particles have factorized scattering,
the LL and RR scattering being given by an $S$ matrix which, as a function of the rapidities,
is the same as the S matrix of the bulk sine-Gordon model, $S_{LL}=S_{RR}=S^{SG}$,
while the LR scattering is trivial. The solitons and antisolitons
 scatter off the boundary one by one 
with no particle production, and the $R$ matrix is given by
\eqn\kondormat{\eqalign{R_\pm^\mp&\equiv R =
-i\tanh\left({\beta-\beta_B\over 2}-{i\pi\over 4}\right)\cr
R_\pm^\pm&=0.\cr}}
In the so called repulsive regime, that is for $g\geq {1\over 2}$, there are
no bound states, and the soliton and antisoliton are the 
only particles in the spectrum. The boundary state can be written in a form similar 
to \bdrstate\
\eqn\kondobdrst{\eqalign{|B\rangle \propto&\sum_{n=0}^\infty \int_{\beta_1<\ldots<\beta_n}
 \prod_i
{d\beta_i\over 2\pi}K(\beta_i-\beta_B)\sum_{\epsilon_i=\pm}
Z_{L,\epsilon_1}^*(\beta_1)\ldots Z_{L,\epsilon_n}^*(\beta_n)\cr
&\times Z_{R,\epsilon_1}^*(\beta_1)\ldots Z_{R,\epsilon_n}^*(\beta_n)|0\rangle,\cr}}
with, as for the Ising case, $K(\beta-\beta_B)=i\tanh{\beta_B-\beta\over 2}$. 
As a result,  introducing ($C$ denotes the chirality, $C=L,R$)
\eqn\newmultiparto{{\cal I}_{2k+1}|\beta_1\ldots\beta_n\rangle_{C_1,\epsilon_1\ldots}=
{\lambda_{2k+1}\over 2}\left(\sum_i e^{(2k+1)\beta_i}\right)|\beta_1\ldots\beta_n
\rangle_{C_1,\epsilon_1\ldots},}
we can write 
\eqn\newmain{|B\rangle=\exp\left
[\sum_{k=0}^\infty {-2\over (2k+1)} e^{-(2k+1)\beta_B}
{{\cal I}_{2k+1}\over \lambda_{2k+1}}\right]|B_{IR}\rangle.}
The coefficients $\lambda_{2k+1}$ 
in \newmultiparto\ will be adjusted for later convenience.

Indeed, a new difficulty arises here
 when one wishes to reexpress the set of commuting
 quantities ${\cal I}_{2k+1}$ 
in terms of local fields. As far as we know, this question
 was first addressed quantitatively  in 
\FS, where the first few conserved quantities
 were studied
numerically using the TBA. The following analytical expression was obtained
in unpublished works by the present authors,  as well as  by Al. Zamolodchikov 
\Zamoprivate, and probably 
by a few others too. A derivation is presented in the appendix for completeness;
to our knowledge, it has never appeared elsewhere, though the technique is hardly 
original.

To proceed, we need to chose some normalizations. 
We first introduce the twisted stress energy tensor
\eqn\stressen{T_{zz}=-2\pi:\left(\partial\phi\right)^2:+i(1-g)\sqrt{2\pi\over g}\partial^2\phi,}
where $\phi\equiv\phi_R$ is the right moving component of the boson. The central charge 
corresponding to this tensor is 
\eqn\central{c=1-6{(1-g)^2\over g}.}
A set of commuting  quantities is then obtained by integrating 
successive powers of this stress energy tensor. We define 
(the $2\pi$ normalization
makes subsequent formulas simpler)
\eqn\fieldsexpres{\eqalign{{\cal O}_2&={1\over 4\pi}\left(T_{zz}+T_{\bar{z}\bar{z}}\right)
\cr
{\cal O}_4&={1\over 4\pi}\left(:T_{zz}^2+R\to L\right) \cr
{\cal O}_6&= {1\over 4\pi}\left(:T_{zz}^3:-{c+2\over 12} :T_{zz}\partial^2T_{zz}:+R\to L
\right)\cr
&\ldots\cr}}
The normalization is such that 
 ${\cal O}_{2k+2}$ goes as ${1\over 2}(-1)^{k+1}(2\pi)^{k}(\partial\phi)^{2k+2}$.
We then define ${\cal I}_{2k+1}=\int_{-\infty}^\infty {\cal O}_{2k+2}dy$.
 Let us stress
that these quantities commute at the conformal point only (in fact, of course,
their left and right components independently commute). In the massive sine-Gordon model
(ie with the bulk  perturbation  $\cos2\sqrt{2\pi g}\Phi$ in our notations), there exists  non chiral
 deformations of these quantities that still commute, and act as sums of odd powers of momenta 
on the (massive) multiparticle states \miscel.
 In the massless scattering description
we are using here, one considers the free boson as the 
limit of the massive sine-Gordon model, and the particular chiral quantities ${\cal I}_{2k+1}$ are
 singled out, which act again as sums of odd powers of momenta on the multiparticle
states. Of course, there are more conserved quantities
right at the conformal point, but they do not seem to have any simple meaning
in terms of rapidities - see next section however, and \FS\ for more 
details.

With this choice, one has (see the appendix)
\eqn\lambdd{\lambda_{2k+1}=\left({\pi\over g}\right)^k(k+1)!
{\Gamma\left[{(2k+1)g\over 2(1-g)}\right]\over
 \left(\Gamma\left[{g\over 2(1-g)}\right]\right)^{2k+1}}
{\left(\Gamma\left[{1\over 2(1-g)}\right]\right)^{2k+1}\over
\Gamma\left[{(2k+1)\over 2(1-g)}\right]}.}
In the following, we will also need 
the relation between the parameter $\beta_B$ of the $R$ matrix
and the coupling $\lambda$ in the action of the Kondo model.
 This was determined in \FLSbig, and 
 reads
\eqn\relat{T_B={\Gamma\left({g\over 2(1-g)}\right)\over
\sqrt{\pi}\Gamma\left({1\over 2(1-g)}\right)} 
\left[\lambda\Gamma(1-g)\right]^{1/(1-g)}.}
{}From \newmain, we then obtain
\eqn\newirexpa{{\cal H}={\cal H}_{IR}+\sum_{k=0}^\infty b_{2k+1}
\lambda^{-{1+2k\over 1-g}}
{\cal O}^o_{2k+2},}
with
\eqn\fromsasha{\eqalign{b_{2k+1}=&{2\over 2k+1}e^{-(2k+1)\beta_B}
{\lambda^{1+2k\over 1-g}\over \lambda_{2k+1}}\cr
=&
\sqrt{\pi}{g^{k+1}\over (1-g)(k+1)!}{\Gamma\left[(k+1/2){1\over 1-g}\right]
\over \Gamma\left[1+(k+1/2){g\over 1-g}\right]}\left[\Gamma(1-g)\right]^
{-{1+2k\over 1-g}},\cr}}
and ${\cal O}_{2k+2}^o$ follows from the expression for ${\cal O}_{2k+2}$ 
by replacing $z$ by $w$, and multiplying by an overall factor $(-1)^{k+1}$:
\eqn\fieldsexprescross{\eqalign{{\cal O}_2^o&=-{1\over 4\pi}\left(T_{ww}+
T_{\bar{w}\bar{w}}\right)
\cr
{\cal O}_4^o&={1\over 4\pi}\left(:T_{ww}^2+R\to L\right) \cr
{\cal O}_6^o&=- {1\over 4\pi}\left(:T_{ww}^3:-{c+2\over 12} :T_{ww}\partial^2T_{ww}:+R\to L
\right)\cr
&\ldots\cr}}

The foregoing results essentially coincide  with those in \BLZ. 
Our route is quite different however; in
particular,  the form of the $R$ matrix or the 
normalization of the  integrals of motion are not used at all in \BLZ, where,
instead, a functional relation approach is developed.

The Kondo model is a very interesting physical example from the point of 
view of the IR perturbation theory. For any value of $g$ (which physically
corresponds to the anisotropy), the IR fixed point is always the same (see \Hew\ for
details and references). To get back
to a $g$ dependent UV fixed point,  one needs to perturb the IR fixed point
by the same family of operators (stress tensor and the like)
but with coefficients that depend on $g$: it is only through this fine tuning
of the coefficients that different flows can be obtained. 
For a given $g$, the free energy for an arbitrary IR perturbation - that expands on the 
conserved quantities  - has an expression similar to what we wrote in the Ising model.

The foregoing  analysis could be generalized to the regime where the 
associated bulk sine-Gordon model
has bound states, that is $g<{1\over 2}$.
The  final expressions involving quantum fields, for instance 
  \newirexpa, would {\sl not} change; they are expected to be analytical in $g$, a result
that can easily be checked using the method we explain below. On the other hand,
expressions 
involving scattering quantities {\sl would} change . Here, we would like to make a remark
 concerning \newmultiparto. 
Because in the scattering the numbers of solitons and breathers are independently
conserved, one expects in general a result of the form, introducing the 
color $\epsilon_i$ for particles ($\epsilon=1,\ldots,m,\ldots$ for breathers, $\epsilon=\pm 1$ 
for solitons antisolitons)
\eqn\newmultipi{{\cal I}_{2k+1}|\beta_1,\ldots,\beta_n\rangle_{C_1,\epsilon_1,\ldots}=
{1\over 2}
\left(\sum_i \lambda_{2k+1,\alpha_i}e^{(2k+1)\beta_i}\right)
|\beta_1,\ldots,\beta_n\rangle_{C_1,\epsilon_1,\ldots},}
where the $\lambda_{2k+1,\alpha}$ are a priori all different. The determination
of these factors is an interesting problem by itself. It can be quickly 
solved if one observes that the formula for the boundary state \kondobdrst\ 
immediately generalizes to the case where breathers are present in the spectrum,
by using the m-breather reflection matrix
\eqn\breat{R_m=-
{\tanh\left({\beta-\beta_B\over 2}-{i\pi\over 4}{mg\over 1-g}\right)
\over \tanh\left({\beta-\beta_B\over 2}+{i\pi\over 4}{mg\over 1-g}\right)}.}
Expanding ${1\over i}{d\over d\beta}\ln R$ in (odd) powers of $e^\beta$, one finds
\eqn\reflctmat{\eqalign{{1\over i}{d\over d\beta}\ln R_m&=4\sum_{k=0}^\infty 
e^{-(2k+1)\beta}\sin\left[ m\pi{(2k+1) g\over 2(1-g)}\right]\cr
{1\over i}{d\over d\beta}\ln R&=2\sum_{k=0}^\infty (-1)^k e^{-(2k+1)\beta}.\cr}}
By putting these expansions in the formula for the boundary state,
it follows that the ratio of normalizations of conserved quantities is the same
as the ratio of the odd powers of $e^\beta$ in \reflctmat, that is
\eqn\ratio{
{\lambda_{2k+1,m}\over \lambda_{2k+1,\pm}}=2 (-1)^k \sin \left[ m\pi{(2k+1)g\over 
2(1-g)}\right].}
The rest of the arguments follows with minor modifications.

The key feature of the spin $1/2$ Kondo problem is that the IR fixed point
is approached along the conserved quantities ${\cal O}^o_{2k+2}$ of even dimensions. The
situation is more interesting  for the higher spin case, or
 the boundary sine-Gordon case.

\subsec{The boundary sine-Gordon problem}

The previous structure generalizes in a slightly more 
complicated form to the boundary sine-gordon model
\eqn\bdrsg{{\cal H}={1\over 2}\int_{-\infty}^0dx\int_{-\infty}^\infty
dy \left[(\partial_x\Phi)^2+(\Pi)^2\right]+2\lambda\cos\sqrt{2\pi g}\Phi(0).}
The boundary interaction is
integrable, and the same manipulations we carried out for the Ising model
can be acomplished here too. While technically  more involved, the general
 spirit
is very similar, so we will restrict ourselves to the salient features.

The quickest way to proceed is to restrict to the attractive case of the associated bulk
sine-Gordon model,
 $g=1/\hbox{integer}$, and to
consider  the reflection matrices
in that case \Gh: 
\eqn\bdrsgrefmat{\eqalign{{1\over i}{d\over d\beta}\ln R_m=2 \sum_{k=0}^\infty
(-1)^k e^{-(2k+1)\beta}{\sin\left[ m\pi (2k+1) g/2(1-g)\right]
\over \sin \left[\pi (2k+1)g/2(1-g)\right]}\cr
{1\over i}{d\over d\beta}\ln (R^+_+\pm R^+_-)=-{1-g\over g}\sum_{k=1}^\infty
(-1)^k e^{-2k\beta(1-g)/g}\tan k\pi {1-g\over g}\cr
\pm {1-g\over g}
\sum_{k=0}^\infty (-1)^{k+1} e^{-(2k+1)\beta(1-g)/g}
+\sum_{k=0}^\infty {e^{-(2k+1)\beta}\over \sin \left[\pi (2k+1) g/2(1-g)\right]}.\cr}}
The boundary scattering is non diagonal in the soliton antisoliton basis,
but it is diagonal for the symmetric and antisymmetric
combinations, which scatter
with the amplitudes $R_+^+\pm R^+_-\equiv R_\pm$. The bulk scattering is diagonal 
in either basis. 

We also recall that the relation between the coupling constant $\lambda$ in the action
and the rapidity $\beta_B$ is modified in the case of the boundary sine-Gordon model,
 reading then
\eqn\newrelation{T_B=\left(2\sin\pi g\right)^{1/(1-g)}
{\Gamma\left({g\over 2(1-g)}\right)\over
\sqrt{\pi}\Gamma\left({1\over 2(1-g)}\right)} 
\left[\lambda\Gamma(1-g)\right]^{1/(1-g)}.}

By using the reflection matrices, and following the same logic as before, we 
obtain immediately the coefficients of all conserved quantities
for the hamiltonian near the IR fixed point. We have now an expansion similar to 
\newirexpa, but with the coefficents $b_{2k+1}$ replaced by  
\eqn\newcoeffs{c_{2k+1}={(-1)^{k}\over 2}{1
\over \sin \left[\pi (2k+1)g/2(1-g)\right]}{1\over \left(2\sin \pi g\right)^{2k+1\over 
2(1-g)}} b_{2k+1},}
where the prefactor is just the ratio of the coefficients of odd powers of $e^{-\beta}$ 
in \bdrsgrefmat\ and \reflctmat\ (of course, this ratio is the same for the breathers
and the soliton antisoliton R matrices), plus an additional power of $2\sin\pi g$ 
arising from the difference between \newrelation\ and \relat.

It is well known  \ABL\ that, at the conformal point, the (chiral part) of the quantities
${\cal I}_{2k+1}$ commute not only together and with the integral
of the perturbation $\int_{-\infty}^\infty dy e^{\pm i\sqrt{8\pi g}\phi}$, but they also commute
with the ``dual'' of the perturbation 
$\int_{-\infty}^\infty e^{\pm i\sqrt{8\pi\over g}\phi}$. 
When the perturbation is  turned on, a deformation of these  
quantities turns out to still be conserved, guaranteeing the integrability of 
the flow. This conservation is true all the way to the IR fixed point,
where again the purely chiral quantities are conserved, by conformal
invariance. It follows that, if one investigates the conservation
perturbatively near the IR fixed point within the dimensionnally regularized scheme,
the only operator that can be added to ${\cal H}_{IR}$, besides the ${\cal O}_{2k+2}$,
is $\cos\sqrt{2\pi\over g}\tilde{\Phi}$. Here, $\tilde{\Phi}$ is the dual
of the field $\Phi$, $\tilde{\Phi}=\phi_R-\phi_L$, and we used that $\phi_R=-\phi_L$ at the
IR fixed point.  By dimensional analysis, its amplitude goes as
 $\lambda^{-1/g}$. The exact amplitude
follows from eqn (6.18) and (6.20) in  \FLSbig. One can thus  finally write 
\eqn\newerirexpa{{\cal H}={\cal H}_{IR}+2\lambda_d \cos\sqrt{2\pi\over g}\tilde{\Phi}
+\sum_{k=0}^\infty c_{2k+1}
\lambda^{-{1+2k\over 1-g}}
{\cal O}_{2k+2}^o,}
where 
\eqn\dualcoupl{\lambda_d={1\over 2\pi g}\Gamma\left({1\over g}\right)
\left[{g\Gamma(g)\over 2\pi}\right]^{1\over g}
\lambda^{-{1\over g}}.}

Observe that the R matrix elements for breathers expand only
on odd powers of $e^\beta$ - this indicates that the non local
conserved quantities formed with $\cos\sqrt{2\pi\over g}\tilde{\Phi}$
have vanishing eigenvalue on the breather 
states, a result of their charge neutrality.

\subsec{The Kondo model with higher spin}

It
is necessary to still generalize the previous arguments slightly, to take into account  the Kondo model
 with higher spin. The structure is actually very similar to the spin $1/2$ Kondo and the BSG case.

The UV Kondo hamiltonian reads as \bdrsg\ with now the spin in a spin $j$ representation
of $U_qsl(2)$, $q=e^{i\pi g}$. One finds that the  hamiltonian near the IR fixed point reads
\eqn\higherspin{{\cal H}={\cal H}_{IR}+\lambda_d\left[S_-e^{i\sqrt{2\pi\over g}
\tilde{\Phi}(0)}
+S_+e^{-i\sqrt{2\pi\over g}\tilde{\Phi}(0)}\right]+\sum_{k=0}^\infty d_{2k+1}
\lambda^{-{1+2k\over 1-g}}{\cal O}_{2k+2}^o,}
where $s$ is in the representation $j-{1\over 2}$, $\lambda_d\propto \lambda^{-1/g}$,
and the coefficients $d_{2k+1}$ could  be determined using the same method as before (see section 5 for more
details) .

\newsec{An application: the resistivity in the Kondo model.}

A good testing ground for the previous considerations is the
isotropic Kondo model, where the strong coupling behaviour
can be probed by experiments at low temperatures. The most 
interesting quantity in that case is of course the resistivity, for which
no closed form results were available so far, besides the $T^2$ term that follows 
from Fermi 
liquid theory \ref\Noz{P. Nozi\`eres, J. Low Temp. Phys. 17 (1974) 31.}
(attempts to compute $\rho$ with the Bethe ansatz have failed,
partly because it is truly a three dimensional quantity). 
The method we have developed in this paper allows us to make an important 
progress on that question: short of getting $\rho$ in closed form, we can at least
compute it perturbatively near the strong coupling fixed point, now that
we know the exact structure of the hamiltonian. This allows us to go beyond the 
Fermi liquid approximation, and evaluate $\rho$ as a power series in $T^2$ 
at low temperatures.

In order to study the resistivity, we first need to go back to the
3d formulation of the system with electronic anihilation operator
\eqn\electd{
\Psi(\vec r)=\int {d^3 \vec  p\over (2\pi)^{3/2}}
e^{i\vec p\cdot \vec r} \Psi(\vec p),
}
where we suppressed the spin indice for simplicity.
As usual,
since the Kondo interaction, when the impurities are dilute, is
assumed to be with only one impurity, we can consider only
the s-wave component of that operator around the Fermi points
\eqn\swaveelec{
\Psi(\vec r)={1\over 2\sqrt{2} i\pi r}[
e^{i k_F r} \psi_R(r)-e^{-ik_F r}\psi_L(r)]
}
with $r>0$ and we used right and left one dimensional moving
field.  This decomposition implies $\psi_L(0)=\psi_R(0)$.
In the interacting theory, only the s-wave  parts of the
three dimensional Green's function will be affected, moreover,
only the $LR$ and $RL$ components of the dimensionally
reduced model are affected by the interaction.
This leads to the following form for the three dimensional
interacting Green's function (for the spin
up field for example)
\eqn\threegreen{\eqalign{
G(\omega_M,&\vec r_1,\vec r_2)-G^0(\omega_M,\vec r_1-\vec r_2)\cr &
={-1\over 8\pi^2 r_1 r_2}\left[ e^{-ik_F (r_1+r_2)}(
G_{LR}(\omega_M,r_1,r_2)-G_{LR}^0(\omega_M,r_1,r_2)) \right.
\cr &\left.
+e^{ik_F(r_1+r_2)}(G_{RL}(\omega_M,r_1,r_2)-G_{RL}^0(\omega_M,r_1,r_2)
)\right],
}}
with the superscript $0$ denoting the free Green function.
As we will see, this is the quantity we need to compute
the resistivity.
The interacting $LR$ (resp. $RL$) Green's
functions are defined by
\eqn\ferdef{
G_{C_1C_2}(\omega_M,r)=-\int_{-\beta/2}^{\beta/2} dy \ e^{i\omega_M y}
\langle \psi_{C_1}(r,y)\psi_{C_2}(0,0)\rangle ,
}
with $C_i$ indicating the chirality.  As an example,
we have\foot{The fermions operators have an extra $\sqrt{2\pi}$
in their normalisation here.}
\eqn\propag{\eqalign{
G_{RL}^0(\omega_M,r) &= \int_{-\beta/2}^{\beta/2} dy
{e^{i\omega_M y}\over
{\beta\over \pi}\sin{\pi\over \beta}(-y+ir)}\cr
&= -2\pi i e^{-\omega_M r} \theta (\omega_M),
}}
where we have used the fact that $r>0$ and that in the UV
\eqn\ferprop{
\langle \psi_R(w_1) \psi_L(\bar{w}_2)\rangle=
-{1\over {\beta\over \pi}\sin{\pi\over \beta}(w_1-\bar{w}_2)}.
}
In the IR, the only difference is the boundary condition
which will result in a change of sign in the propagator.
When we put everything back into the three dimensional
expression for the Green function, we get (at the IR fixed
point)
\eqn\threegreen{\eqalign{
&G^{IR}(\omega_M,\vec r_1,\vec r_2)-G^0(\omega_M,\vec r_1-\vec r_2)\cr &
={i\over 2\pi r_1 r_2}\left[ e^{-ik_F (r_1+r_2)}
  e^{\omega_M (r_1+r_2)}
\theta(-\omega_M)
-e^{ik_F(r_1+r_2)} e^{-\omega_M (r_1+r_2)} \theta(\omega_M)
\right]\cr &
=G^0(\omega_M, \vec r_1) T(\omega_M) G^0(\omega_M,-\vec r_2).
}}
Following the arguments of \affleck, for a dilute array of
impurities of densities $n_i$ the lowest order correction
to the complete Green function takes the form
\eqn\loweord{
G(\omega_M,\vec r_1,\vec r_2)-G^0(\omega_M,\vec r_1-\vec r_2)
\simeq n_i \int d^3 \vec r_i
G^0(\omega_M,\vec r_1-\vec r_i)T(\omega_M)
G^0(\omega_M,\vec r_i-\vec r_2)
}
Summing over multi-impurity terms, the self-energy takes
the simple form
\eqn\self{
\Sigma (\omega_M)=n_i T(\omega_M)
}
where higher orders in $n_i$ are neglected.  The retarded
self-energy is found by the analytical continuation
$i\omega_M\rightarrow \omega+i\eta$ leading to
\eqn\selfr{
\Sigma^R(\omega)=-{i n_i \over \pi \nu}
}
$\nu$ is the number of spin per channel, we have
reestablished its dependance at the end since it only amounts to
a factor of two (separate spins contribute the same).
This is the expected result at the IR fixed point for the
one channel Kondo model.  Finally the resistivity follows
from the Kubo formula for the conductivity
\eqn\kubo{
{1\over \rho(T)}=\sigma(T)=2{e^2\over 3 m^2}\int
{d^3\vec p\over (2\pi)^3} \left[
-{dn\over d\epsilon_k}\right]
\vec p \cdot \vec p \ \tau(\epsilon_k),
}
with the single particle lifetime defined by
$1/\tau=-2 Im \Sigma^R$.  The dispersion relation
$\epsilon_k=v_F k$ has been linearised in that limit.

All this discussion was done using the fermions but to continue
and understand how to get away from the IR fixed point, we need
to use our earlier results.
To make contact with our previous discussion of the Kondo model,
we need to bosonise the system.  This is done using the rules
\eqn\bosoferm{
\psi_{L/R,\mu}(r,y)\propto e^{\pm i\sqrt{4\pi}\phi_{L/R,\mu}(r,y)}.
}
Notice that we have reestablished the spin dependence,
$\mu=\uparrow,\downarrow$, since this will be crucial
in the following.
At the UV fixed point we have $\psi_{L,\mu}(0)=\psi_{R,\mu}(0)$ but since
we are interested rather in perturbation around the IR
fixed point, we impose the conditions
$\psi_{R,\mu}(0)=-\psi_{L,\mu}(0)$ for the IR correlators.
This leads to the RL (LR) bosonic propagator
\eqn\propag{
\langle \phi_{R,\mu}(w_1)\phi_{L,\nu} (\bar{w}_2)\rangle=
\delta_{\mu\nu}\left[-{1\over 4\pi}\ln {\beta\over \pi}\sin
{\pi\over \beta}(w_1-\bar{w}_2)\right]
}
which translates in the correct fermionic propagator when
using the bosonisation rules given above.
Although we are interested in computing the Green
function of spin up fields, for example, the integrable
description is rather in terms
of the spin and charge densities, ie
introduce
\eqn\spincharge{\eqalign{
\phi_s&={1\over \sqrt{2}}(\phi_\uparrow-\phi_\downarrow)\cr
\phi_c&={1\over \sqrt{2}}(\phi_\uparrow+\phi_\downarrow).
}}
In terms of these fields, the interaction at the boundary
only involve the spin field and is given by the hamiltonian
written in the previous section.  The charge field remains
non-interacting.
The perturbation around the IR fixed point is described  by the
hamiltonian
\eqn\perthalm{
{\cal H}={\cal H}_{IR}+
\sum_{k=0}^\infty b_{2k+1} \lambda^{-{1+2k\over 1-g}}
{\cal O}_{2k+2}^o
}
where all the couplings and operators have been given in section 3;
the boson field in the latter section coincides with  $\phi_s$ here.
On the other hand, if we look at the bosonisation of the
Green function for the spin up field, we observe that
there will be contributions for each field
\eqn\contrib{
\psi_{L\uparrow}\propto e^{i\sqrt{4\pi}\phi_{L\uparrow}}
=e^{i\sqrt{2\pi}(\phi_{Lc}+\phi_{Ls})}
}
and when computing the interacting left-right Green
function for example, the charge sector will be completely
decoupled, ie
$$
\langle \cdots \rangle= \langle \cdots \rangle_{charge}
\times \langle \cdots \rangle_{spin}
$$
Only when doing the Fourier transform will
the charge part contribute.   Let us proceed to the computation
to show this more explicitly.

The isotropic case ($g=1$) leads to some simplifications
in the previous expressions, leading to the identification
\eqn\pertcoup{
b_{2k+1} \lambda^{-{(1+2k)\over (1-g)}}=
{1\over \pi^k (k+{1\over 2}) (k+1)!} T_B^{-(1+2k)}.
}
The coupling $T_B$ now is identified with the usual
Kondo temperature $T_K$ (up to a normalization that is a matter of convention,
and will be decided later) and the contribution of each operator
is determined through these relations.
This provides the information necessary to compute higher
corrections to the resistivity from the IR fixed point.

Up to order $T_B^{-2}$ the contributions
are exactly the same
as the ones found previously since only one operator, the
energy momentum tensor, appears to that order.
It is at the third order that the non-trivial approach to
the fixed point will be needed since the second operator
${\cal O}_{4}$ will be involved.  First let us proceed
to reproduce results found before for
the two first orders using our bosonised formulation.
To first order, the leading irrelevant operator is (with the proper
normalisation)
\eqn\firstc{\eqalign{
-{1\over 4\pi}(T_{ww}+T_{\bar{w}\bar{w}})&=
{1\over 2}[:(\partial_w\phi_s)^2:+:(\partial_{\bar{w}}\phi_s)^2:]
-{(2\pi T^2)\over 24}
\cr &=
:(\partial_y\phi_s)^2:-{(2\pi T^2)\over 24}
}}
where we have used the fact that the operator is inserted
at $r=0$ to get the last line.  The constant is a disconnected term
that gets cancelled when dividing by the partition function to evaluate the
correlator: we can thus forget about it in what follows.
Inserting \firstc\ in the correlator (of the relevant
RL or LR components) we get the lowest order
contribution to the one dimensional propagators (again for the
spin up field for example)
\eqn\lowestord{\eqalign{
{2\over T_B}\int_{-\beta/2}^{\beta/2} dy dy' \ e^{i\omega_M y}
\langle e^{\pm i\sqrt{2\pi}[\phi_{R/L,c}+\phi_{R/L,s}](r_1,y)}
:(\partial_{y'}\phi_s)^2:\times\cr
\times e^{\mp i \sqrt{2\pi}[\phi_{L/R,c}+\phi_{L/R,s}](r_2,0)}\rangle_{IR}\cr}
}
with the subscript IR meaning that we evaluate the propagators
with respect to the IR action.  Note that the contribution
from the charge boson decouples and the perturbation only
affects the spin sector.
Again let us write explicitely
the $RL$ component: we have for the first correction
\eqn\lowestordex{\eqalign{
\delta^{(1)} G_{RL}&={-1\over 4\pi T_B}\int_{-\beta/2}^{\beta/2} dy dy'
\ e^{i\omega_M y}
{{\beta\over \pi}\sin{\pi\over \beta}(w_1-\bar{w}_2)\over
[{\beta\over \pi}\sin{\pi\over \beta}(w'-\bar{w}_2)]^2
[{\beta\over \pi}\sin{\pi\over \beta}(w'-w_1)]^2} \cr &
=-{2i\pi\over T_B} \epsilon(\omega_M) \omega_M e^{-\omega_M (r_1+r_2)},
}}
with $\epsilon(\omega_M)$ the step function.
This leads to a correction of the self-energy of the form
\eqn\correcfirst{
\Sigma^R(\omega)=-{in_i\over 2\pi \nu}\left[2+i{\omega\over T_B}\right]
}
which is the expected form.
The correction is real and does not
contribute to the conductivity or the life time.  To get bona-fide
contributions, we need to go further in the IR perturbation theory.
To next order, the conserved quantity ${\cal O}_2$ will contribute
again but the higher quantity ${\cal O}_4$ will not yet give
a contribution.
So to second order, we have
\eqn\higherord{\eqalign{
-{2\over T_B^2}&\int_{-\beta/2}^{\beta/2} dy dy'dy'' \ {e^{i\omega_M y}
\over [{\beta\over \pi}\sin {\pi\over \beta}(w_1-\bar{w}_2)]^{1/2}}
\times
\cr & \times
\langle e^{i\sqrt{2\pi}\phi_{R,s}(w_1)}
:(\partial_{y'}\phi_s)^2:
:(\partial_{y''}\phi_s)^2:
e^{- i \sqrt{2\pi}\phi_{L,s}(\bar{w}_2)}\rangle_{IR}
}}
where we already contracted the charge part.
Using the relation
\eqn\usefulrel{\eqalign{
:(&\partial_{y'}\phi(0,y'))^2::(\partial_{y''}\phi(0,y''))^2:
=:(\partial_{y'}\phi(0,y'))^2 (\partial_{y''}\phi(0,y''))^2:
\cr &+4\left( {-1\over 4\pi [{\beta\over \pi}\sin
{\pi\over \beta}(y'-y'')]^2}\right)
:(\partial_{y'}\phi(0,y')) (\partial_{y''}\phi(0,y'')):\cr
& +2 \left( {-1\over 4\pi [{\beta\over \pi}
\sin{\pi\over \beta}(y'-y'')]^2}\right)^2,
}}
we get three contributions to the second order, two of which
are divergent.  The regularisation of divergences here is
done by analyticity, as explained in section  3:
we slightly modify
the contours of the $y''$ integral, and move it by $i\delta$
in the complex plane.  The integrals are then done by simple
residue evaluation.  Usually there could be a dependance on
the way the contour is deformed but this disapears here since
the operators commute with each other (there is no simple
pole in their OPE).
The last term in the expansion has no
frequency dependence and the explicit evaluation (using our
prescription for the regularisation of the divergence) gives
zero.  The first contribution has the form
\eqn\firstcont{\eqalign{
\delta^{(2a)}G_{RL}=&-{2\over T_B^2 (8\pi)^2}\int_{-\beta/2}^{\beta/2}
dy dy' dy'' e^{i\omega_M y} \
[{\beta\over\pi}\sin{\pi\over\beta}(w_1-\bar{w}_2)]^3\times
\cr 
&\times\left\{ {1\over
[{\beta\over\pi}\sin{\pi\over\beta}(w'-w_1)]
[{\beta\over\pi}\sin{\pi\over\beta}(w'-\bar{w}_2)]
}\right\}^2\times\cr
&\times
\left\{ {1\over
[{\beta\over\pi}\sin{\pi\over\beta}(w''-w_1)]
[{\beta\over\pi}\sin{\pi\over\beta}(w''-\bar{w}_2)]
}\right\}^2\cr
}}
which contains no divergences and can be evaluated
straightforwardly by the method of residues.  The integral over
$y',y''$ leads to
\eqn\partistep{
\delta^{(2a)}G_{RL}={1\over 2 T_B^2}
\int_{-\beta/2}^{\beta/2}
dy e^{i\omega_M y} {[\cos{\pi\over\beta}(w_1-\bar{w}_2)]^2
\over [{\beta\over\pi}\sin{\pi\over\beta}(w_1-\bar{w}_2)]^3},
}
and evaluation of the integral gives
\eqn\evalfirst{
\delta^{(2a)}G_{RL}={i\pi\over 2 T_B^2}
\epsilon(\omega_M)e^{-\omega_M(r_1+r_2)}
[\omega_M^2+(\pi T)^2].
}
The second contribution, which has divergences, takes the form
\eqn\seclast{\eqalign{
&\delta^{(2b)}G_{RL}={-1\over 4 \pi^2 T_B^2}\int_{-\beta/2}^{\beta/2}
dy dy' dy'' e^{i\omega_M y} \ {1\over
[{\beta\over\pi}\sin{\pi\over \beta}(y'-y'')]^2}
\times
\cr 
&\times {{\beta\over\pi}\sin{\pi\over \beta}(w_1-
\bar{w}_2)\over
[{\beta\over\pi}\sin{\pi\over \beta}(w'-\bar{w}_2)]
[{\beta\over\pi}\sin{\pi\over \beta}(w'-w_1)]
[{\beta\over\pi}\sin{\pi\over \beta}(w''-\bar{w}_2)]
[{\beta\over\pi}\sin{\pi\over \beta}(w''-w_1)]}
}}
and evaluating by residues leads to
\eqn\resdeux{
\delta^{(2b)}G_{RL}={i\pi\over 2 T_B^2}
\epsilon(\omega_M) e^{-\omega_M (r_1+r_2)}
(2\omega_M^2 -2(\pi T)^2).
}
So the total contribution to second order to the retarded
green's function takes the form
(once we analytically continue to real frequencies)
\eqn\realsec{
\Sigma^R(\omega)==-{in_i\over 2\pi \nu}\left[2+i{\omega\over T_B}
-  {1\over 4 T_B^2} (3\omega^2+(\pi T)^2)\right]
}
and as expected we have a universal function of
$(\omega/T_K, T/T_K)$.
The previous results did not require any information about
the other operators but at third order, the operator
\eqn\ofour{\eqalign{
{\cal O}_4&=-{1\over 4\pi} [:T_{ww}^2:+:T_{\bar{w}\bar{w}}^2:]
\cr 
&=\pi\left[
:(\partial_y\phi_s)^4:-{1\over 2\pi}:\partial_y\phi_s\partial^3_y\phi_s:
\right] -{(\pi T)^2\over 2}:(\partial_y\phi_s)^2:+
{3 (\pi T)^4 \over 80\pi}
}}
needs to be taken into accout: it comes with the coupling
$1/(3\pi T_B^3)$ in the hamiltonian. Using the relation
\eqn\useful{\eqalign{
:(\partial_{y'}\phi)^n: \ :e^{-i\sqrt{2\pi}\phi_L(\bar{w}_2)}:=&
\sum_{p=0}^n {n\choose p}
\left( {i\over \sqrt{8\pi}[{\beta\over\pi}\sin
{\pi\over\beta} (w'-\bar{w}_2)]}\right)^{n-p}\times\cr
&\times
:(\partial_{y'}\phi)^p e^{-i\sqrt{2\pi}
\phi_L(\bar{w}_2)}:\cr}
}
we get, using the residue theorem, the contribution,
\eqn\necorr{
\delta^{(3a)}G_{RL}=
{1\over 24 T_B^3}
\epsilon(\omega_M) e^{-\omega_M(r_1+r_2)} [
6 (i \omega_M)^3+6 i\omega_M (\pi T)^2].
}
There is also a contribution from the leading irrelevant operator
when expanded to third order, which reads
\eqn\necorri{
\delta^{(3b)} G_{RL}={\pi\over 6 T_B^3} \epsilon(\omega_M)
e^{-\omega_M (r_1+r_2)}[3 i\omega_M (\pi T)^2+5 (i\omega_M)^3]
}
At this order, the contributions are
all imaginary and we need to go the the next order to get non trivial
contributions to the resistivity.
At fourth order there are
two contributions, one coming from the leading operator only,
${\cal O}_2^4$, and
another, from the combination of the leading and next to leading
operators, ${\cal O}_2 {\cal O}_4$.
The computation are analogous to the previous ones,
but  more tedious.  The final result for the retarded self
energy up to fourth order is
\eqn\realseci{\eqalign{
\Sigma^R(\omega)=&-{in_i\over 2\pi \nu}\left[2+i{\omega\over T_B}
-  {1\over 4 T_B^2} \left(3\omega^2+\left(\pi T\right)^2\right)-\right.\cr &
-i\left({5\over 12}+{3\over 24\pi}\right)\left({\omega\over T_B}\right)^3-i
\left({1\over 4}+{1\over 8\pi}\right){\omega\over T_B}\left({\pi T\over T_B}
\right)^2+\cr &
\left.+\left({35\over 192}+{7\over 32\pi}\right)\left({\omega\over T_B}\right)^4+
\left({19\over 96}+{5\over 16\pi}\right)\left({\pi T\over T_B}\right)^2\left({\omega\over 
T_B}\right)^2+\right.\cr &
\left. +\left({11\over 192}+{3\over 32 \pi}\right)\left({\pi T\over T_B}\right)^4\right]
}}
Using the Kubo formula this leads to our main result for the
resistivity (which we computed to sixth order)
\eqn\resisiii{\eqalign{
\rho(T)=&{3 n_i\over (\pi v_F \nu)^2}
\left[
1-{1\over 4}\left({\pi T\over T_B}\right)^2+\left({13\over 240}
+{3\over 20\pi}\right) \left({\pi T\over T_B}\right)^4 \right.
\cr & \left. +\left(
{47\over 10080}-{1\over 8\pi}-{53\over 336\pi^2}\right)
\left({\pi T\over T_B}\right)^6  \right]
}}
In the following figure we compare this result with the numerical
renormalisation group method \cost.  The definition of $T_B$
is related to the usual Kondo temperature through a simple
factor $T_B={2\over \pi} T_K$.
\fig{Comparison with Numerical results.}{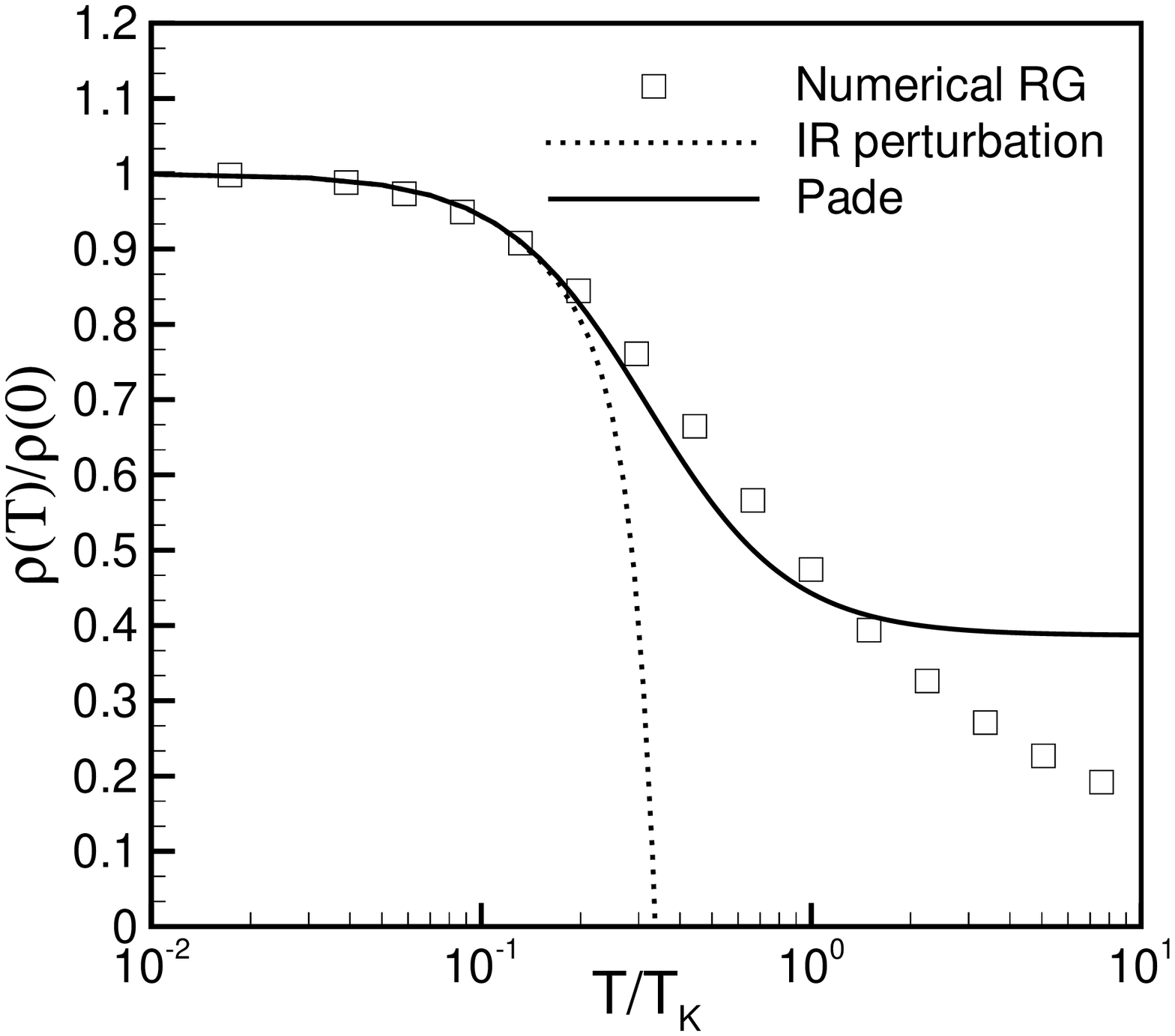}{8cm}
\figlabel\tabb
The agreement is quite good considering that there is no fitting
parameter. The Pad\'e approximants were found to be very stable, and
give a control of the curve $\rho(T)$  all the way to $T\approx T_K$,
which is right in the crossover region. It is thus clear that our method provides a 
good analytical understanding of the strong coupling resistivity.

\newsec{Another application: duality}

The general structure of the IR hamiltonians is given by a set of local conserved quantities,
plus at most one non local conserved quantity. This implies some duality properties 
that we now discuss. 

\subsec{Duality in Kondo with higher spin}

The main thing about expression \higherspin\ is that
it contains only {\sl one} type of exponential \foot{Here, we should 
stress that this is
a regularization dependent feature, that holds for our dimensionnally regularized approach. Ohter
exponentials would appear as counter terms in other approaches.}.
Qualitatively, this is a consequence of integrability: it is natural 
to expect the trajectory to appear integrable both from the UV and IR
fixed point; on the other hand, theories with several harmonics are
generally non integrable - therefore, only one harmonic can occur. Quantitatively,
this  leads to a very strong similarity of  the physical properties
expanded near the UV or near the IR fixed point, after 
replacement $g\to {1\over g}$;
in particular, quantities that are ``blind'' to the integer spin conserved quantities, if any, 
will exhibit a complete {\sl duality} symmetry 
between the UV and the IR.

To discuss the matter further, let us
 compute the boundary free energy at vanishing temperature and with an applied 
field $2H S_z$ ($S_z$ taking values $j,j-1,\ldots, -j$ in the representation of 
spin $j$). We introduce 
the quantity $\epsilon_{2j}$ defined by 
\eqn\defeps{\tilde{\epsilon}_{2j}=\int {d\omega\over 2\pi}{\sinh\left[
\left({1\over 1-g}-n\right){\pi\omega\over 2}\right]\over\sinh\left[
{g\over 1-g}{\pi\omega\over 2}\right]}
\tilde{\epsilon}(\omega)-2V(2j-1).}
Here $\epsilon$ is the quantity defined in  eq. (6.9) of \FLSbig\ with the conventions of the appendix ($M=2,\hbar=1,V\equiv 2V,e=1$); one has 
in particular, $\epsilon_1=\epsilon$ of \FLSbig. 
The parameter $V$ is related with the field by $H=gV$.
One can then establish,
from the well known TBA formula in the limit $T\to 0$, 
that (this generalizes slightly \TW. See also \pauldual,\usdual)
\eqn\kondofreeen{\eqalign{f=&\int {d\beta\over 2\pi}{1\over \cosh(\beta-\beta_B)}\epsilon_{2j}(\beta)\cr
=&V\int {d\omega\over 2\pi} e^{i\omega(A-\beta_B)}
{1\over 2\cosh{\pi\omega\over 2}}
{\sinh\left({1\over 1-g}-2j\right){\pi\omega\over 2}\over
\sinh{g\over 1-g}{\pi\omega\over 2}}
{G_-(\omega)G_+(0)\over \omega(\omega-i)}-V(2j-1).}}
In this formula, 
\eqn\gggdef{G_-(\omega)=\sqrt{2\pi\over g} {\Gamma\left[i\omega/2(1-g)\right]
\over \Gamma\left[i\omega g/2(1-g)\right]\Gamma\left[1/2+i\omega/2\right]}e^{i\omega\Delta}
,}
and $\Delta={1\over 2}\ln{1-g\over g}+{1\over 2(1-g)}\ln g$. To compute $f$, we close the contour in the upper half plane when $A>\beta_B$. The only poles
 are those at $\omega=2(1-g)ni$, $n$ a positive integer. The UV  expansion of $f$ follows
\eqn\uvexp{\eqalign{f=&V\sqrt{\pi}\sum_{n=1}^\infty {(-1)^{2nj+n}\over n n!}
{\sin 2jn\pi g\over \sin 2n\pi g}{e^{-2(1-g)n\Delta}\over \Gamma\left(-ng\right) 
\Gamma\left[3/2-n(1-g)\right]}\left({T_B\over e^A}\right)^{2n(1-g)}\cr
&-2jVg.\cr}}
We now recall the  correspondence between the cut-off $A$ and the physical field in 
that case $e^A=V{G_+(0)\over G_+(i)}$. Using that the field coupled to the impurity in 
the Kondo problem is $H=gV$, together with the correspondence between the bare coupling $\lambda$ 
and $T_B$, one has
\eqn\morecorresp{{e^A\over T_B}={H e^\Delta\over \left[\lambda \Gamma(1-g)\right]^{1/1-g}}.}
This allow us to rewrite the free energy in the form
\eqn\kondofuv{\eqalign{f=&{\sqrt{\pi}H\over g}\sum_{n=1}^\infty {(-1)^{2nj+n}\over n n!}
{\sin 2jn\pi g\over \sin 2n\pi g}{1\over \Gamma\left(-ng\right) 
\Gamma\left[3/2-n(1-g)\right]}\left[{\lambda\Gamma(1-g)\over H^{1-g}}\right]^{2n}\cr
&-2jH.\cr}}

When $A<\beta_B$ on the other hand, we close the contour in the lower half plane. There are 
now two types of poles: the ones at $\omega=-(2n+1)i$ give contributions to the free energy of the 
form $\left({e^A\over T_B}\right)^{2n+1}$, while those of the form $\omega=-2ni{1-g\over g}$ give
 the contribution (which we will refer to as ``non-analytic'')
\eqn\irexp{\eqalign{f_{non-analytic}=&\sqrt{\pi} H\sum_{n=1}^\infty
{(-1)^{2nj}\over n n!}{\sin  (2j-1)n\pi/g\over\sin 2n\pi/g}
{1\over \Gamma\left(-n/g\right)\Gamma\left[3/2-n(1-1/g)\right]}
\cr
&\left[{\lambda\Gamma(1-g)\over H^{1-g}}\right]^{-{2n\over g}}
-(2j-1){H\over g}.\cr}}
{}From this we deduce the relation  (to be used in \higherspin)
\eqn\newdualcoupl{\lambda_d={\sin{\pi\over g}\over \pi g}\Gamma\left({1\over g}\right)
\left[{\sin\pi g\over \pi}g\Gamma(g)\right]^{1\over g}\lambda^{-{1\over g}},}
together\foot{Observe that \newdualcoupl\ is very similar
 to \dualcoupl. It would become identical if the
 the boundary sine-Gordon term came with the coupling $2\lambda\sin \pi g$, which 
is actually the natural choice
within the quantum group framework underlying these problems.} with 
\eqn\dualkondo{f\left(j,\lambda,H,g\right)\equiv f\left(j-{1\over 2},\lambda_d,{H\over g},{1\over g}\right),}
where the equality holds up to analytical terms (odd powers) in $H/T_B$. 

This duality has an obvious physical origin. We can compute the free energy near the UV fixed point
perturbatively in powers of $\lambda$, or near the IR fixed point
perturbatively in powers of $\lambda_d\propto\lambda^{-1/g}$ and in powers of $\lambda^{-1/(1-g)}$. 
The firs type of terms comes from the Kondo type interaction near the IR 
fixed point, that looks formally like the one near the UV fixed point, but
with the replacements  $j\to j-1/2$, $g\to 1/g$  and $H\to H/g$. It is interesting 
to discuss the later replacement in more details - the physical interaction near the UV and IR 
fixed points of course does not change, it is always  $2HS_z$. However, to take this into account in
the integrable approach, one needs to trade this term for a shift of the field $\phi$ 
 in the Kondo interaction: the way this trading takes place depends on the charge 
of the exponentials, and this is why there is a rescaling in the TBA expressions, which are formally
computed using an action with a $H$ dependent Kondo coupling:  see \FL\ and below 
for
 more details. Since all the 
integrals near the IR fixed point are defined by analytical continuation, they clearly lead to 
results obeying \dualkondo. In addition, the local and non local integrals of motion commute: 
therefore, in physical properties that involve the logarithm
of the partition function (for instance), the terms coming from the Kondo type 
perturbation near the IR fixed point {\sl do not} mix with the terms coming 
from the integer spin conserved quantities \foot{This remarkable property
is clearly visible on the logarithms of the reflectiom matrix elements; see 
the formula \bdrsgrefmat\ for a completely analogous example 
in the context of the boundary sine-Gordon model.}. Therefore, to the non analytic contribution to $f$ near 
the IR fixed point, is simply {\sl added} an analytic contribution in odd powers of $H/T_B$.
The structure of this analytic contribution is actually extremely simple, and depends only
weakly on the spin.

Of course, the argument establishing the duality also holds at non vanishing
temperature. Even though no close expression is known for the 
free energy in that case, we thus expect \dualkondo\ to still hold, this time up to 
terms analytical in powers of $H/T_B,T/T_B$. 

\subsec{Duality in the boundary sine-Gordon model}

That  \newerirexpa\  contains only {\sl one} cosine  has a simple physical
meaning here -  the flow approaches the IR fixed point 
along a direction where there is  a term in the hamiltonian
corresponding to  tunneling of electrons, but no term for  tunneling of pairs, triplets etc.

The most interesting properties to study in that context are transport properties,
for which a non equilibrium formalsim such as Keldysh is required. Some of our
conventions are discussed in the appendix; here, we will concentrate on the salient
features only. Introducing a vector potential $a(t)$, the current is computed
by $I(t)={\delta\ln Z\over \delta a}$, and expanded perturbatively near the UV or IR fixed point.
Consider the UV fixed point first: there, the potential vector can be reabsorbed into
the cosine term by a shift of the boson, so, restricting to constant voltage $V$, the current 
expands as a series of Coulomb gas integrals somewhat similar to the ones in equilibrium; the key difference
however, is that the vertex operators $V_\pm=\exp\pm i\sqrt{2\pi g}\Phi$ 
are integrated on the Keldysh contour, represented in figure 3, and that contour ordered propagators
are used. 
\fig{Keldysh contour.}{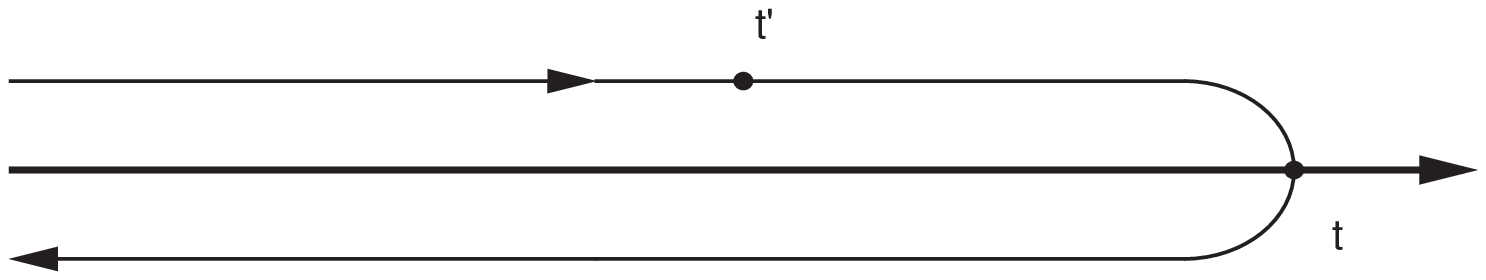}{6.5cm}
\figlabel\tabb
More specifically, a vertex operator stands 
at the extremity of the contour at time $t$, while  other operators are integrated on the contour. The
only non vanishing contributions are the ones which are electrically neutral. At non vanishing  voltage 
$V$, 
each vertex operator $V_\pm(t')$ comes with an additional phase  $e^{i\pm gVt'}$. The contour ordered propagator
is such that $\langle T_c \left[\Phi(t')\Phi(t'')\right]\rangle = -{1\over 2\pi}\ln (t_>-t_<)$,
where $t_>$ denotes the time that is the latest as measured along the contour, and $t_<$ 
the time that is the earliest. For instance, for $t'$ on the above or lower part of the 
contour as in figure 3, the contraction that appears in the computation of the current is, $\langle T_c\left[e^{i\sqrt{2\pi g}\Phi(t)}e^{-i\sqrt{2\pi g}\Phi(t')}\right]
\rangle={1\over (t-t')^{2g}},
\hbox{resp. }{1\over  (t'-t)^{2g}}$. The non trivial  monodromy  of the vertex operators
ensures that the contribution of the two parts of the contour do not cancel out, and a non trivial 
result is obtained (see the appendix for some examples). At finite temperature $T$, 
the only change is that, in  the propagator, $\ln t$ is replaced by $\ln {\sinh \pi Tt\over \pi T}$.

 Now, our point is not so much to 
discuss the structure of this expansion (many details on this issue
can be found in \Claudio\ for instance),
but to comment on the duality properties it might give rise to. For this, let us investigate
the  computation of the current in the IR. If, in our framework of dimensional 
regularization, the only operator in the IR were the $\cos\sqrt{2\pi\over g}\tilde{\Phi}$, duality would 
easily follow from the matching of the two expansions. The complication we have to discuss
is the role of all the ${\cal O}_{2k+2}$ operators added to the hamiltonian. 

Let us first make a crucial observation. Consider for instance the contour ordered propagator
in the two situations of figure 4, 
\fig{Conserved quantity on the contour.}{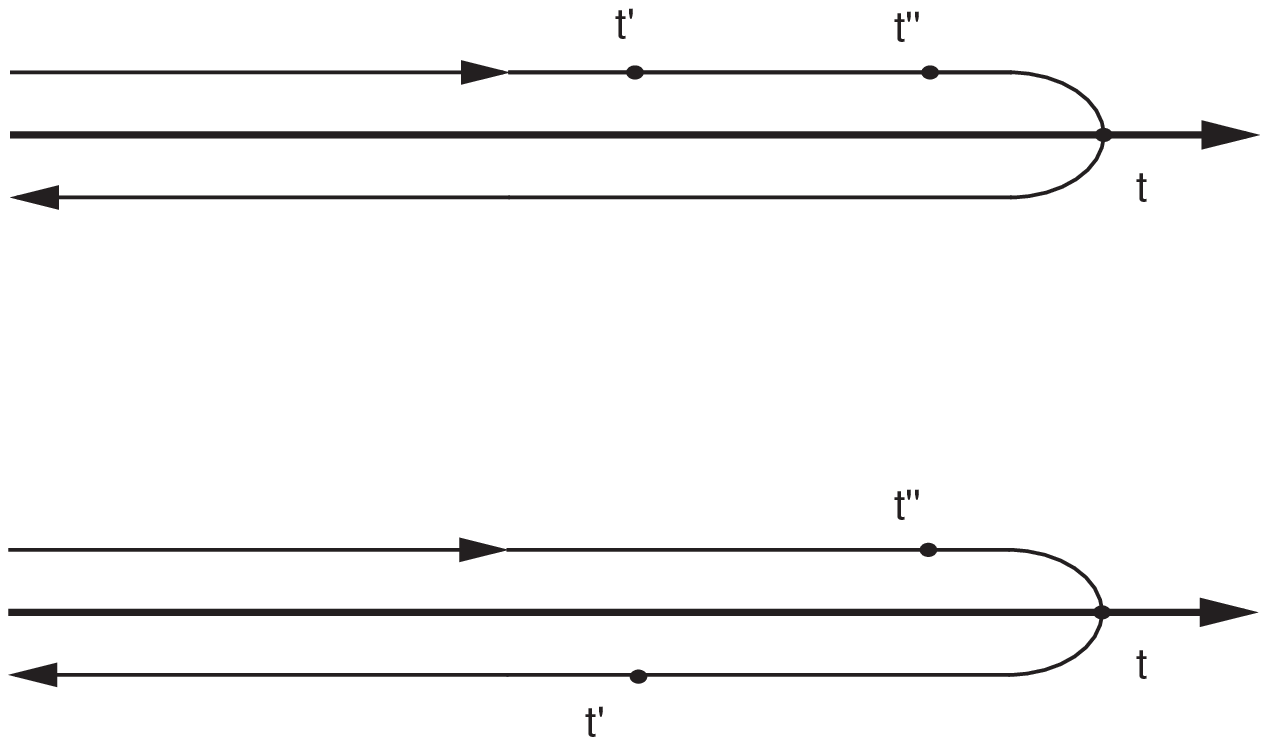}{6.5cm}
\figlabel\tabb
where say $\partial\Phi(t')$  (we call in this paragraph $\Phi$ what
is in fact the dual of the original field $\tilde{\Phi}$ for notational simplicity; $\partial$ denotes 
time derivative)
is inserted at time $t'$, 
and some expression $X(t'')$ at time $t''$. In the first situation where $t'$ occurs earlier 
on the contour, 
$$
\langle T_c\left[\partial\Phi(t') X(t'')\right]\rangle=\sum_{contractions} {1\over 2\pi(t''-t')}\hat{X}(t''),
$$
where $\hat{X}$ denotes the remainder in $X$ once contracted. In the second situation where
$t'$ occurs later on the contour,
$$
\langle T_c\left[\partial\Phi(t') X(t'')\right]\rangle=\sum_{contractions} {-1\over 2\pi(t'-t'')}\hat{X}(t''),
$$
and of course, the two expressions are actually equal. This easily generalizes to cases where $\partial\Phi$
 is replaced
by any polynomial in derivatives of $\Phi$, in particular the ${\cal O}_{2k+2}$. As for $X$,
it can be one of the ${\cal O}_{2k+2}$ itself, as well as a product of such an operator by a vertex operator,
the result still holds: in other words, the result of the contraction does not depend on the 
order on the contour, and for the ${\cal O}_{2k+2}$ operators, integrals along the Keldysh contour
just behave like ordinary contour integrals - a somewhat trivial fact, once one remembers that
there is no cut in the complex plane for the contractions involved here.

This observation being made, we observe that the ${\cal O}_{2k+2}$ in the IR hamiltonian
give rise to two complications. First, when one trades the coupling of the vector potential $\int a\partial_t\Phi$
for a shift of the field $\Phi$, this time not only does one get a shift ${1\over g}Vt$ in the argument 
of the cosine; one also  gets in the new hamiltonian additional terms made of $a$ and polynomials 
in the derivatives of $\Phi$ (for instance, the $:\left(\partial\Phi\right)^4:$ in $:T^2:$
gives rise to a $a^2:\left(\partial\Phi\right)^2:$ term, etc). Thus, when defining the current
as  the functional derivative of $\ln Z$ with respect to $a$, one gets, in the IR, a more complicated 
expression
than in the UV: what has to be inserted at $t$ on the Keldysh contour is the sum of a vertex operator 
and a series of polynomials in derivatives of $\Phi$. 

Now consider the perturbative computation of this current in the IR: we have to insert on the Keldysh 
contour either vertex operators or operators ${\cal O}_{2k+2}$. For the component of the current at $t$
that is not the vertex operator however, no cut is necessary at $t$. According to the  
observation above, the integrals on the Keldysh contour of the various insertions
then just behave like ordinary integrals, for which the upper and lower parts of the contour 
cancel out - in other words, the current is still obtained  by only inserting vertex operators at $t$,
in complete analogy with the UV case.

The second complication due to the ${\cal O}_{2k+2}$ is that these operators contribute to the 
perturbation series in the IR. Consider thus a generic term in the perturbation series,
where a few vertex operators as well as conserved quantities have been inserted. 
To regulate divergences, it might be necessary to slightly displace the contours - this does not
matter anyway, as we now argue. Indeed, consider moving the contours for the insertions of conserved quantities,
say ${\cal O}_{2k+2}$ and ${\cal O}_{2k'+2}$. Since they are polynomials in derivatives of $\Phi$, according to 
our  observation above, these contours can be deformed as for usual
integrals.  The residue of their short distance expansion is 
a total derivative, so when we move one contour through the other, we are left with 
the contour integral of a total derivative. If in turn we try to deform this contour to zero, since the short
distance expansion of a total derivative with any quantity cannot have a  simple pole, no obstacle
is met. In other words, we can freely pass through one another the contours for conserved quantities
${\cal O}_{2k+2}$. 

Let us now try to pass these contours through the vertex operators. Consider thus a situation as the one in
figure 5 where we have four vertex operators inserted on the Keldysh contour, and are trying 
to pass the ${\cal O}_{2k+2}$ contour through them. 
\fig{Vertex operators.}{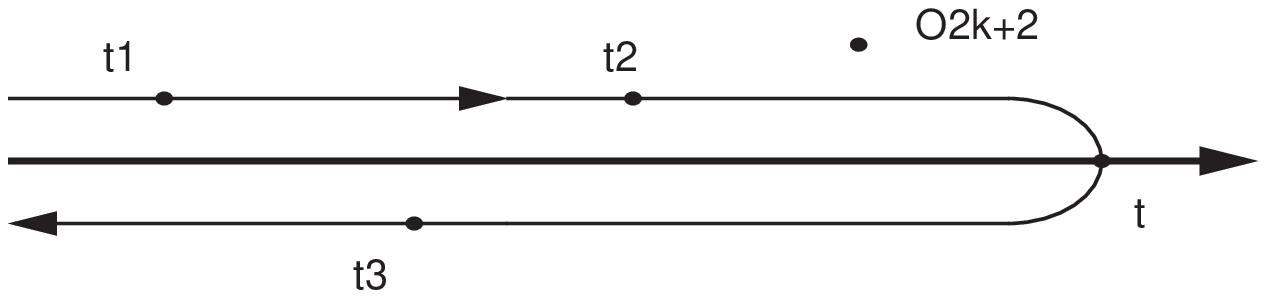}{6.5cm}
\figlabel\tabb
In doing so, we encouter four poles, whose residues
are total derivatives. Let us call the residue of the expansion of ${\cal O}_{2k+2}$ and $V_\epsilon$,
$\partial {\cal O}_{k,\epsilon}$. If $t_1,t_2,t_3,t$ are the arguments of the four vertex operators,
the total quantity picked up is 
$$
\partial_{t_1} {\cal O}_{k,\epsilon_1}V_{\epsilon_2}(t_2)V_{\epsilon_3}(t_3)V_{\epsilon}(t) 
+\hbox{ permutations}
$$
Instead of  contour integrating this quantity, let us simply look at its contour ordered average. Because
the various contractions depend only on the {\sl difference} of arguments, the effect of 
 summing over permutations is to compute the derivatives with respect of sums of arguments
of quantities that depend only on their differences, that is, is zero. Hence, the contribution of the 
residues when moving the ${\cal O}_{2k+2}$ contour through the vertex insertions cancels out,
and we can squeeze this contour to zero. In other words, the ${\cal O}_{2k+2}$ {\sl do not}
contribute to the current at all. This is independent of the voltage or the temperature. As far as the current
goes therefore, it is fully determined, in the scheme where integrals
are analytically regularized, by the $2\lambda\cos\sqrt{2\pi g}\Phi$ perturbation in the UV 
 and $2\lambda_d\cos\sqrt{2\pi\over g}\tilde{\Phi}$ in the IR. This allows us 
to conclude that
\eqn\duality{I(\lambda,g,V,T)=gV-gI\left(\lambda_d,{1\over g},gV,T\right).}

\newsec{Conclusions}

In conclusion, we would like to stress that the implementation of the IR perturbation
theory, as well as the existence of duality, rely completely on the integrability
of the problem.  The latter
acts as a symmetry that restricts the IR hamiltonian in a  drastic fashion, so that the structure 
of the IR perturbation is almost the same as the UV one, maybe up to analytical terms. In general 
impurity problems, we do not expect the duality to be more than a quick qualitative argument to find the 
leading irrelevant operator. We also do not expect IR perturbation theory to make much sense,
because of the difficulty in regularizing higher order terms when the operators 
do not commute. 

\bigskip
\noindent{\sl Acknowledgments:}  We  thank I. Affleck,
C. de C. Chamon, T.A. Costi, P. Fendley, 
E. Fradkin, D. Freed, A.C. Hewson and S. Zamolodchikov 
for many illuminating discussions. We also thank 
P. Fendley for a related collaboration. This work was supported
by the DOE and the NSF (under the NYI program).

\appendix{A}{Normalization of conserved quantities in the sine-Gordon model}

We discuss here the problem of determining the constant $\lambda_{2k+1}$ in the definition 
\newmultiparto. To do so, we consider the free action in the bulk,
to which we add two perturbations: one of them is a field coupled to 
the $U(1)$ charge, and the other a term proportional to 
the conserved quantity ${\cal I}_{2k+1}$. Going to a hamiltonian description 
in the closed string  channel, we obtain
\eqn\testact{{\cal H}={\cal H}_{free}+ V{\sqrt{2g\over\pi }}\int_{-\infty}^\infty dy
 \partial_y
\Phi+\mu\int_{-\infty}^\infty   {\cal O}_{2k+2}.}
The question we then consider is the ground state energy of this theory. 
It can be computed using the integrable structure \HMN,\Zamo,\FLSbig. We use
 here the notations of section VI of the
latter reference; in addition, we set $M=2$, $\hbar=1$, $V\equiv 2V$ and $e=1$. 
We denote the rapidity by $\beta$ instead of $\theta$.  The 
constant $\lambda$ of this reference corresponds, in the present
paper, to $\lambda\equiv{1\over g}-1$; it of course has nothing to do with
what we call $\lambda$ in the present paper, that is the bare coupling. 

The problem factorizes into R and L components. Consider say the R sector. 
The equations
determining the ground state density of particles 
depend only on the momenta, which are not affected by the perturbations
of the hamiltonian. The cut-off $A$ however changes, in a way that depends on the 
perturbations in a crucial way. The equivalent of eqn (6.9) of \FLSbig\ 
is now
\eqn\epsilquan{V-e^\beta-\mu {\lambda_{2k+1}\over 2}e^{(2k+1)\beta}=
\epsilon(\beta)-\int_{-\infty}^A \Phi(\beta-\beta')\epsilon(\beta')d\beta'.}
It follows, if the Fourier transforms are defined as in \FLSbig, and if we denote
 $\epsilon_-(\omega)=\tilde{\omega}e^{-i\omega A}$, that
\eqn\newepsfour{\eqalign{\epsilon_-(\omega)=-{1\over i}{G_-(\omega)G_+(i)\over \omega-i}
e^A+{V\over i}{G_-(\omega)G_+(0)\over\omega}\cr
-{\mu\lambda_{2k+1}\over 2i} 
{G_-(\omega)G_+[(2k+1)i]\over \omega-(2k+1)i}e^{(2k+1)A},}}
where the kernels are given in \gggdef\ above, and 
in eqn (6.6) of \FLSbig. The cutoff $A$ is such that $\epsilon(A)=0$ ie $\lim_{\omega\to\infty}
\omega\epsilon_-(i\omega)=0$. We will restrict ourselves to the 
case where $V,\mu>>1$, where the first term in \newepsfour\ becomes
negligible. It then follows that
\eqn\cutoff{e^{(2k+1)A}={2V\over \mu\lambda_{2k+1}}{G_+(0)\over G_+[(2k+1)i]}.}
We can then compute the energy per unit length
\eqn\energ{E=2\int_{-\infty}^A d\beta\rho(\beta)\left[e^\beta+\mu\lambda_{2k+1}
e^{(2k+1)\beta}-V\right]\approx 2\mu\lambda_{2k+1}-2V\tilde{\rho}(0)
\tilde{\rho}[-(2k+1)i],}
where again we neglected the first term, of order one. Using another result 
form  \FLSbig\
\eqn\densfou{\tilde{\rho}(\omega)={1\over 2i\pi}{G_-(\omega)G_+(i)\over \omega-i}
e^{(i\omega+1)A},}
we get, after some algebra,
\eqn\efinal{E=-{1\over \pi}{2k+1\over 2k+2} V^{2k+2\over 2k+1}
\left({2\over \mu\lambda_{2k+1}}\right)^{1\over 2k+1}
{G_+(i)G_+(0)^{2k+2\over 2k+1}\over G_+[(2k+1)i]^{1\over 2k+1}}.}
On the other hand, we can compute the energy directly 
from the hamiltonian. In the integral of  ${\cal O}_{2k+2}$, only
the leading term contributes since all the others involve second or 
higher derivatives
of $\phi$, which vanish at the saddle point. Therefore one has, using the normalization
in \fieldsexpres,
\eqn\directen{E=-2V\sqrt{2g\over\pi}  {2k+1\over 2k+2}
V^{2k+2\over 2k+1}\left({2\over \mu}\right)^{1\over 2k+1} 
\left({\sqrt{2g\over\pi}\over 2k+2}\right)^{1\over 2k+1}
{1\over (-2\pi)^{k\over 2k+1}}.}
{}From this, it finally follows that
\eqn\finalresult{\lambda_{2k+1}=\left({\pi\over g}\right)^k (k+1)!
\left({\Gamma\left[{1\over 2(1-g)}\right]\over\Gamma\left[{g\over 
2(1-g)}\right]}\right)^{2k+1}
{\Gamma\left[{(2k+1)g\over 2(1-g)}\right]\over \Gamma\left[{2k+1\over 
2(1-g)}\right]}.}

\appendix{B}{Some remarks on Keldysh and analytic continuation}

In \FLeS, 
a formula for the current was proposed
\eqn\conducconjec{I=gV-ig\pi T{\partial\over \partial \ln\lambda}\ln{Z(p,\lambda)\over Z(-p,\lambda)}.}
Here, $Z(p)$ is an analytic continuation of the partition function at ``imaginary voltage'',
accomplished through a Jack polynomials expansion \FLeS. 

The partition function at imaginary voltage is defined as $\hbox{Tr} e^{-{\cal H}(p)/T},$
where $p$ is an integer and  ${\cal H}(p)$ is obtained from ${\cal H}$ by shifting the argument of the 
exponential $\cos\sqrt{2\pi g}\Phi\to \cos\left(\sqrt{2\pi g}\Phi+2\pi pTy\right)$. The physical 
voltage is such that $2\pi pT=igV$, so an analytical continuation in $p$ from integer to imaginary values
has to be carried out. The Keldysh formalism actually tells us how to perform
this continuation. To see this, let us first
recall some basic results. We 
consider the boundary sine-Gordon model with a vector potential $a(y)$
\KF. After the usual shift,
one can write the partition function
\eqn\keldyi{Z=Z_0\left\{1+\lambda^2\int_0^{1/T}\int_0^{1/T} dy_1dy_2\cos[a(y_1)-a(y_2)] p(|y_1-y_2|)+\ldots
\right\},}
where dots stand for higher order terms, we restrict  to $g<{1\over 2}$ so the integrals
are all convergent,  and 
\eqn\keldyii{p(y)=\left({\pi T\over \sin \pi Ty}\right)^{2g}.}
The current then follows from $I={\delta\ln Z\over \delta a}$, together with the usual contour deformation
\eqn\keldyiii{\eqalign{I(t)&=2\lambda^2\int_C {dt'\over i}\sin[a(t)-a(t')]\langle T_c e^{i\sqrt{2\pi g}
\Phi(t)}
e^{-i\sqrt{2\pi g}\Phi(t')}\rangle+\ldots\cr
&=2\lambda^2\int_{-\infty}^t dt'\sin[a(t)-a(t')]{P^>(t-t')-P^<(t-t')\over i}+\ldots,\cr}}
where $P^>(resp. P^<)$ is the analytic continuation to $y=it(resp. y=-it)$ of $P$. 
Let us now restrict to a DC voltage $a(t)=gVt$; in that case,
\eqn\keldyiv{I=\lambda^2 {P(gV)-P(-gV)\over i},}
where
\eqn\keldyv{P(x)=\int_0^\infty dt e^{ixt}{P^>(t)-P^<(t)\over i}.}
One has 
$$
P(x)=2(\pi T)^{2g-1}\sin\pi g \int_0^\infty e^{ixt/\pi T}{dt\over (\sinh t)^{2g}},
$$
The latter integral is tabulated, so one gets
\eqn\keldyvi{P(gV)=(2\pi)^{2g}T^{2g-1}{\sin \pi g\over \sin\left(\pi g-{igV\over 2 T}\right)}
{\Gamma(1-2g)\over \Gamma\left(1-g+{igV\over 2\pi T}\right)\Gamma\left(1-g-{igV\over 2\pi T}\right)}.}

Let us now get back to the question of analytically continuing $Z(p)$ in \conducconjec. 
Consider the case $p$ is an integer, so the quantities $Z(p)$ are well defined.
Of course, for $p$ integer, $Z$ is {\sl even} in $p$, so the argument of the derivative 
in \conducconjec\ is identically zero. Let us nevertheless expand it formally 
 in powers of $\lambda$.  
At lowest order, the contribution to the current involves the quantities
\eqn\keldyvii{Q(p)=\int_0^{1/T}\cos(2\pi pTy)p(y)dy.}
One has
$$
Q(p)=(\pi T)^{2g-1}\int_0^\pi e^{2i py}{dy\over (\sin y)^{2g}}
$$
This integral is tabulated, and, for $p$ an integer, reads
\eqn\interm{Q(p)=(2\pi)^{2g}T^{2g-1}(-1)^p{\Gamma(1-2g)\over \Gamma(1-g+p)\Gamma(1-g-p)}.}
Still for $p$ an integer, the expressions \keldyvi\ and \interm\ coincide: this is because, in that case,
the voltage plays the role of a Matsubara frequency, so what we have here is the 
standard identity between Fourier transforms of temperature Green functions and retarded 
time-Green functions \AGD. But we see now 
how to perform the continuation of  $Z(p)$: we first need to deform the contour, at each order in 
peturbation theory, from  the imaginary time
interval to the Keldysh contour - this is possible for $p$ an integer -  and {\sl then} replace
 $p$ by $igVT/2\pi$ in the integrand.  

At the level of final expressions, that is  \keldyvi\ and \interm, however, it is less clear
what must be done. The ``recipe'' proposed in \FLeS\ consists in expanding the integral $Q$ 
into a sum of rational functions of Gamma functions 
\eqn\jackdef{Q(p)=(2\pi)^{2g}T^{2g-1}\sum_{n=0}^\infty
{\Gamma(g+n) \Gamma(g+
n+p)\over \Gamma^2(g)\Gamma(1+n)
\Gamma(1+n+p)},}
and then perform the continuation simply by replacing $p$ by the appropriate non integer
value in each of the Gamma functions. The sum \jackdef\ was studied in details in \FLeS, where
 it was shown that, for arbitrary $p$, this continuation of $Q$ coincides with \keldyvi.
Therefore, this gives the same result as the one obtained by deforming contour, which is the 
correct physical one, based on the Keldysh analysis. 

Notice that the continuation using the 
expression \jackdef\  is not  the same as the continuation discussed in 
\sergei. In the latter work, the author expands the 
partition functions $Z(p)$ in \conducconjec\ over Matsubara propagators, and then
performs the continuation in $p$ in each term independently. There is no reason why
this definition should coincide with ours, and herefore it is not surprising that 
disagreements are found in \sergei:
only 
{\sl one} continuation is expected to work, and the conjecture made in \FLeS\ is that, for the 
tunneling problem described by  the boundary sine-Gordon model,  it is the one
using the expansions \jackdef, and for higher order, the corresponding 
sums based on Jack polynomials theory (see below). 

To understand better why this might be true, we observe that $P(gV)$ 
has a simple power law behaviour $P\propto (V/T)^{2g-1}$ as $V/T\to\infty$, while $Q(p)$
in \interm\ 
does not. The power law behaviour is expected from the Keldysh contour representation,
and on physical grounds as well: it is necessary for the current to have a finite expansion
in terms of $V/T_B$ as $T\to 0$. On the contrary \jackdef, supplemented by the naive replacement
of $p$ by ${igV\over 2\pi T}$ does have the right behaviour.

When one considers higher powers of $\lambda$ in the expansion of the current using the Keldysh 
formalism, one gets  integrals which still coincide, for $p$ integer, with the integrals
occuring in $Z(p)$, through contour deformation. The challenge, if one whishes  to prove the 
 conjecture  \conducconjec, is to show that replacing $p$ by $igVT/2\pi$ in the 
 deformed contour integrals {\sl coincides} with replacing $p$ by $igVT/2\pi$
in the  expansion
\eqn\keldyshx{\eqalign{Q_{2n}(p)=(2\pi)^{2g}T^{2ng-2n+1} {1\over \Gamma(g)^{2n}}
\sum_{\bf m}
&\prod_{i=1}^n {\Gamma\left[m_i+g(n-i+1)\right]
\over\Gamma\left[m_i+1+g(n-i)\right]}\times\cr
&\times{\Gamma\left[p+m_i+g(n-i+1)\right]\over\Gamma\left[p+m_i+1+g(n-i)\right]},\cr}
}
(where the sum is over all sets ${\bf m}=(m_1,\ldots,m_n)$ with $m_1\geq m_2\ldots\geq m_N\geq 0$), 
a result we explicitely checked above at lowest order. Clearly, the two procedures define analytical 
continuations of functions defined for integers. Under reasonable assumptions, it is known that
two functions that coincide on integers and have the same behaviour at infinity
are actually identical. Therefore, a proof of \conducconjec\ would simply be that
the continuation of \keldyshx\
behaves, when $p\to\pm i\infty$, as $p^{2ng-2n+1}$. Though we have not proven this analytically, we have
checked it numerically for the first few  values of $n$.

\listrefs
\bye